\documentclass[draft]{agujournal2019}

\usepackage{url} 
\usepackage{lineno}
\usepackage{amsmath}
\usepackage{amsfonts}
\usepackage{gensymb}
\usepackage{lineno}
\usepackage{graphicx}
\usepackage{float}
\usepackage{ragged2e}
\usepackage[inline]{trackchanges} 
\usepackage{soul}
\draftfalse

\journalname{Nature Machine Intelligence}

\begin{document}
%
%

\title{Physically Constrained Generative Adversarial Networks for Improving Precipitation Fields from Earth System Models}
\justifying

%
%




\authors{Philipp Hess\affil{1,2}, Markus Drüke\affil{2}, Stefan Petri\affil{2}, Felix M. Strnad\affil{2,3}, and Niklas Boers\affil{1,2,4}}

\affiliation{1}{\small Technical University Munich, Munich, Germany; School of Engineering \& Design, Earth System Modelling}
\affiliation{2}{Potsdam Institute for Climate Impact Research, Member of the Leibniz Association, Potsdam, Germany}
\affiliation{3}{Cluster of Excellence - Machine Learning for Science, Eberhard Karls Universität Tübingen, Germany}
\affiliation{4}{Global Systems Institute and Department of Mathematics, University of Exeter, Exeter, UK}




\correspondingauthor{Philipp Hess}{philipp.hess@tum.de}




\begin{keypoints}
\item A generative adversarial network improves both distributions and spatial structure of the precipitation output of a numerical Earth system model.
\item Constraining its architecture enables the network to generalize to transient future climates not seen during training.
\item A gradient-based interpretability method shows that the network has learned to identify geographical regions with strong model biases. 
\end{keypoints}

%
%

%
%


\begin{abstract}
Precipitation results from complex processes across many scales, making its accurate simulation in Earth system models (ESMs) challenging. Existing post-processing methods can improve ESM simulations locally, but cannot correct errors in modelled spatial patterns.  Here we propose a framework based on physically constrained generative adversarial networks (GANs) to improve local distributions and spatial structure simultaneously. We apply our approach to the computationally efficient ESM CM2Mc-LPJmL. Our method outperforms existing ones in correcting local distributions, and leads to strongly improved spatial patterns especially regarding the intermittency of daily precipitation. Notably, a double-peaked Intertropical Convergence Zone, a common problem in ESMs, is removed. Enforcing a physical constraint to preserve global precipitation sums, the GAN can generalize to future climate scenarios unseen during training. Feature attribution shows that the GAN identifies regions where the ESM exhibits strong biases. Our method constitutes a general framework for correcting ESM variables and enables realistic simulations at a fraction of the computational costs.

\end{abstract}


\section{Introduction}
    
    Numerical Earth system models (ESMs) simulate the dynamics of Earth system components such as the atmosphere, oceans,
    vegetation, and polar ice-sheets, as well as their interactions, by solving the relevant partial differential equations on discretized spatial grids. The grid resolution is limited by computational costs. For state-of-the-art comprehensive ESMs, integrating the differential equations requires parallelized runs on thousands of CPU cores.
    The finite resolution requires processes on unresolved spatial scales to be parameterized, i.e., to be written as functions of the resolved variables. This introduces a source for potential errors in ESMs.
    It is generally expected that the accuracy of ESM simulations increases with increasing resolution of the spatial grid on which the model is integrated \cite{palmer2019scientific}.
    
    A higher grid resolution, however, comes at even higher computational cost, and trade-offs are therefore typically necessary. The time current state-of-the-art ESMs take to make projections for the decadal to centennial time scales relevant in the context of anthropogenic climate change render it challenging to simulate ensembles with sufficient size for a thorough uncertainty quantification. Similarly, the high computational cost even for simulating single trajectories prevent more systematic parameter calibration. Complementary to high-resolution but computationally demanding ESMs, efficient model setups that are still as accurate as possible are therefore also needed.
    
    The generation of precipitation involves a wide range of physical processes, from microscopic interactions of droplets in clouds over atmospheric convection to synoptic-scale weather systems.
    The resulting complex dynamics needs to be captured accurately to model the high variability and intermittency of precipitation in both space and time.
    A reduced resolution and limited number of explicitly resolved processes in ESMs therefore leads to errors that can strongly affect the representation of sub-grid scale processes such as precipitation \cite{wilcox2007frequency, boyle2010impact, ipcc-ar6-wg1-2021}.
    
    These errors can be addressed in a local or point-wise manner by applying post-processing methods to the individual simulated time series.
    Traditionally, this is done by relating the statistics of a historical model simulation with observations.
    Quantile mapping (QM), in particular, has become a popular method for improving the model output statistics of precipitation \cite{deque2007frequency, tong2021bias, gudmundsson2012downscaling, cannon2015bias}.
    It approximates a mapping from the estimated cumulative distribution function of the modelled to the observed quantity over a historical period.
    The inferred mapping can then be applied to correct new data.
    QM gives good results in correcting temporal distributions locally, i.e., errors in the distribution at a given grid cell. QM is, however, not able to improve the spatial structure of the modelled output, such as its intermittency especially for the case of precipitation. For this task a spatial context larger than the single grid cells used to compute the distributions in QM is required. It should be emphasized that even a (almost) perfect reproduction of the distributions at each grid cell would by no means guarantee that also the spatial patterns are reproduced accurately. In particular, the patterns may still be too smooth and lack the spatial intermittency that is typical for realistic precipitation fields.
    
    Machine learning (ML) methods from image-to-image translation in computer vision offer a new approach to improve the structure of ESM output in the spatial dimension.
    Recently, artificial neural networks have been applied successfully to post-processing tasks of numerical weather prediction and climate models \cite{rasp2018neural, gronquist2021deep, franccois2021adjusting}.
    In weather forecasting, the trajectories of the observed state and the numerical weather model starting at an initial condition taken from observations can be directly and quantitatively compared.
    This allows to train discriminative ML models such as deep neural networks \cite{lecun2015deep} to directly minimize a pixel-wise distance measure as a regression task. 
    
    For ESMs tasked with climate projections, such a pixel-wise ground truth is not available, rendering a direct comparison between observed and modelled trajectories impossible. In particular, ML models cannot be trained via minimizing differences between simulations and corresponding observations in this case. The goal of ESMs is indeed to produce long-term summary statistics rather than to agree with observations on short time scales.
    In this context, generative adversarial networks (GANs) \cite{goodfellow2014generative, mirza2014conditional, Isola2017} have emerged as suitable ML models. GANs learn to approximate a target distribution from which realistic samples can be drawn. 
    Crucially, recent developments have shown successful application of cycle-consistent GANs \cite{zhu2017unpaired, Yi_2017_ICCV, hoffman2018cycada} to training tasks that do not require pairwise training samples. This suggests the suitability of cycle-consistent GANs for post-processing Earth system model simulations, for which no direct observational counterpart exists.
    By learning stochastic functions, GANs can also model the small-scale variability that cannot be predicted deterministically.
    This enables them to overcome the problem of blurring that is often found in neural network predictions \cite{ravuri2021skillful}.
    Based on these properties, GANs have been proposed for sub-grid scale parameterizations \cite{gagne2020machine} and statistical downscaling of numerical weather forecasts \cite{price2022increasing,harris2022generative}. Employing GANs in a post-processing task of a regional climate model, \citeA{franccois2021adjusting} found a comparable bias correction skill of their GAN compared to quantile mapping.
    
    Training ML algorithms typically requires the training data and separate test sets for predictions to be independent and identically distributed. 
    When applied to historical observations and transient ESM time series with changing forcing, however, the underlying distributions are non-stationary, i.e., training and test distributions are different. In particular in the context of anthropogenic climate change, this has made the application of ML methods challenging. 
    To generalize to such out-of-sample predictions, physics-informed or constrained neural networks have been proposed. 
    These methods incorporate physical knowledge into the neural network through penalties in the loss function \cite{raissi_physics-informed_2019-2}, or include additional layers \cite{beucler2021enforcing} in the architecture.
    
\begin{figure}[h]
            \centering
            \includegraphics[width=\textwidth]{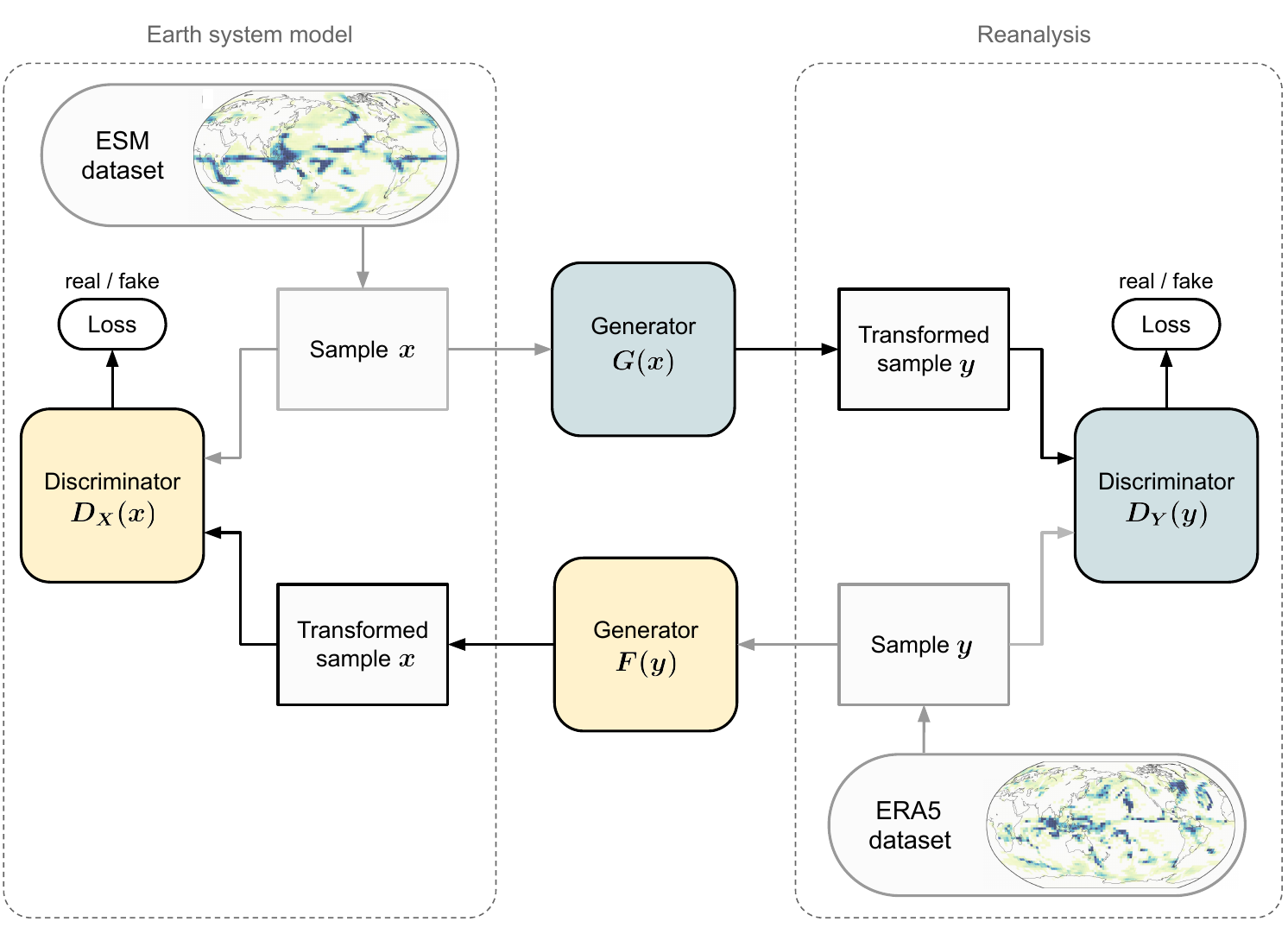}
            \caption{Schematic of the CycleGAN model, showing the two generator-discriminator pairs that learn to translate samples from the ESM simulations to the ERA5 reanalysis (grey) and vice versa (yellow). 
            Training the two generators to learn inverse mappings of each other allows to enforce cycle-consistency in the translation of the unpaired samples, i.e. $x \rightarrow G(x) \rightarrow F(G(x)) \rightarrow \tilde{x} \approx x$ and vice versa for $y$.
            As described by \citeA{zhu2017unpaired}, the cycle-consistency loss (Eq.~\ref{eq:cycle_loss}) is motivated from natural language translation, where one should arrive at the same sentence after translating it into another language and back.
            In the training context, this has been found to improve the stability and to prevent typical problems in adversarial networks, such as mode collapse, where every input would be mapped to the same output image \cite{zhu2017unpaired}.
            }
            \label{fig:gan_setup}
        \end{figure}

    Here, we introduce a physically constrained GAN (see Fig.~\ref{fig:gan_setup} and Methods for details) to improve the precipitation output of ESMs, and demonstrate its performance by applying it to the CM2Mc-LPJmL model \cite{druke2021cm2mc}. 
    We frame the post-processing as an image-to-image translation task with unpaired training samples.
    The first image domain corresponds to the ESM simulations, and the second to daily precipitation fields from the ERA5 reanalysis ``ground truth'' \cite{hersbach2020era5}, spanning the period between 1950 and 2014.
    The translation is performed with a CycleGAN \cite{zhu2017unpaired}, consisting of two generator-discriminator pairs, that learn bijective mappings between the ESM and reanalysis domains, with consistent translation cycles.
    We add a physical constraint as an additional layer to the generator network architecture after training in order to preserve the global precipitation sum (see Methods).
    
    We compare our results to QM-based post-processing as well as the output of a considerably more complex and higher-resolution, state-of-the-art ESM from Phase 6 of the Coupled Model Intercomparison Project (CMIP6), namely the GFDL-ESM4 \cite{gfdlesm4} model.
    Further, the ability of the GAN to generalize to transient future climate scenarios is evaluated for physically constrained and unconstrained GAN architectures.
    When applying neural network models to future projections that cannot (yet) be verified, transparency of the method becomes important. 
    Therefore, we examine whether the GAN's feature attribution is physically reasonable, using the SmoothGrad \cite{smilkov2017smoothgrad} interpretability method (Methods).
    Moreover, the quantitative interpretation of the GAN results allows us to identify regions with particularly large biases of the underlying process-based ESM, which will in turn be helpful for improving its representation of relevant physical mechanisms. For a more detailed description of the methods applied in this study we refer to the Methods section below.
    
\section{Results}

    \subsection{Correcting temporal distributions}
    
\begin{figure}[htp!]
            \centering
            \includegraphics[width=\textwidth]{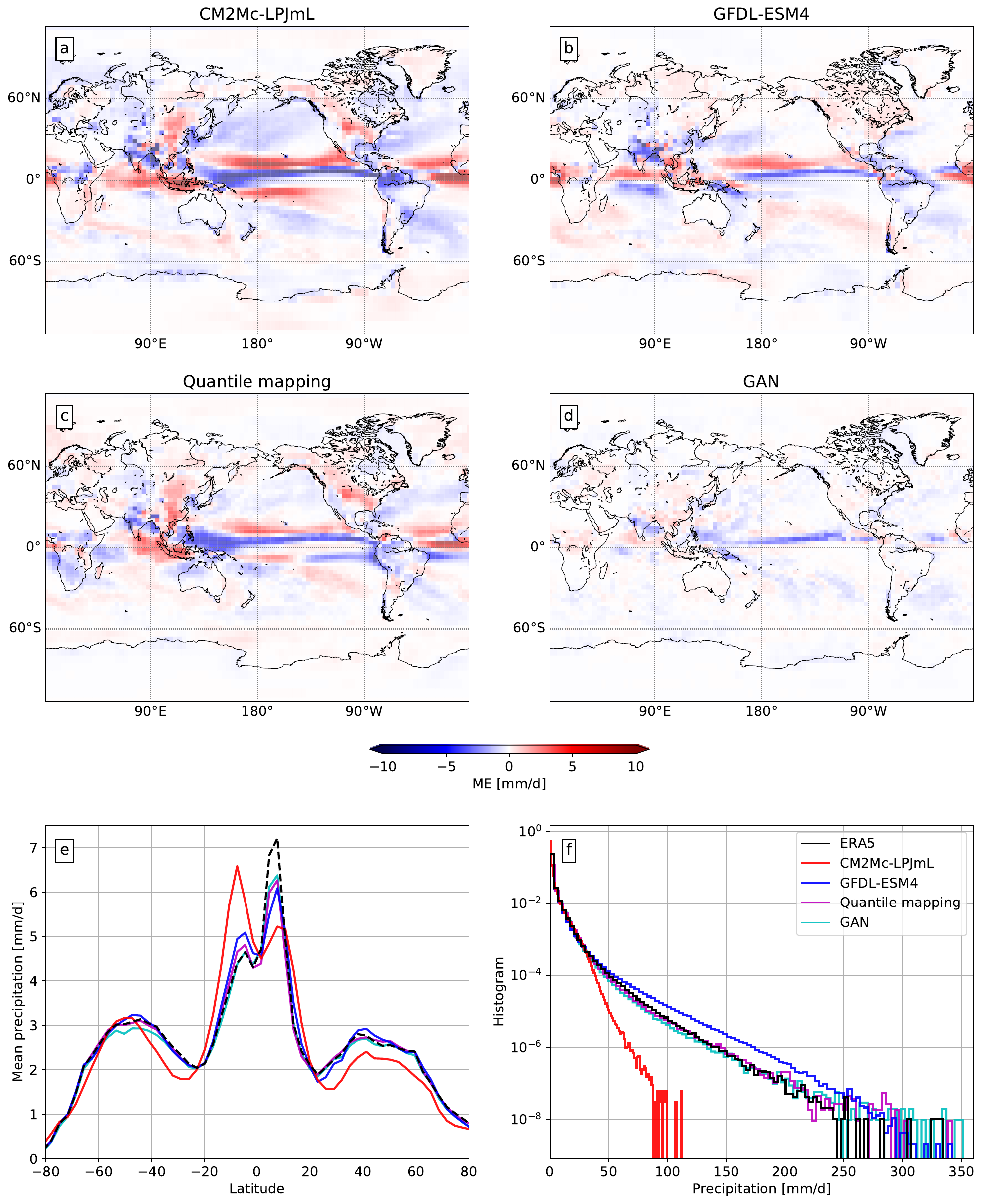}
            \caption{Comparison of global mean error maps over the JJA season, long-term precipitation statistics based on latitude-profiles and relative frequency histograms.
            Mean errors of (a) CM2Mc-LPJmL, (b) GFDL-ESM4, (c) QM-based and (d) GAN-based post-processing methods applied to the CM2Mc-LPJmL output.
            The mean error is computed with respect to the ERA5 reanalysis data. 
            The largest errors are in the tropics, where also the largest mean precipitation values are observed (see panel (e)). 
            The GAN shows the largest error reduction, strongly reducing the double-peaked ITCZ in the tropics. 
            Quantile mapping, on the other hand, is not able to remove the ITCZ bias.
            See Figs.~S1--S4 for corresponding figures for annual time series, as well as the other three seasons. 
            (e) Precipitation rates averaged over time and longitudes and relative frequency histograms (f) are shown for ERA5 data (black), CM2Mc-LPJmL (red), GFDL-ESM4 (blue), quantile mapping (magenta) and the GAN (cyan). The GAN applied to the CM2Mc-LPJmL output corrects the double-peaked ITCZ as well as the histogram over the entire range of precipitation rates.
            }
            \label{fig:temporal_bias}
        \end{figure}

        When comparing the spatial precipitation fields from CM2Mc-LPJmL with the ERA5 data, large biases are evident, especially in the tropics, where a pronounced double-peaked Intertropical Convergence Zone of CM2Mc-LPJmL can be seen (Fig.~\ref{fig:temporal_bias}a).
        The more complex and higher-resolution -- yet computationally much more expensive -- GFDL-ESM4 model exhibits a similar spatial pattern of bias, although with a reduced southern peak (Fig.~\ref{fig:temporal_bias}b). 
        
        We evaluate our method against quantile mapping, which a state-of-the-art method to correct temporal distributions (Fig.~\ref{fig:temporal_bias}c).
        The GAN shows a strongly improved skill overall, and especially in correcting the double-peaked ITCZ (Fig.~\ref{fig:temporal_bias}d), compared to quantile mapping, but also compared to GFDL-ESM4 model. 
     
        This is also summarized in the averaged absolute value
        of the mean error (ME) shown in the spatial plots (Table~\ref{tab:bias}). Here, the GAN shows the strongest error reduction compared to QM and GFDL-ESM4, reducing the error of CM2Mc-LPJmL by 75\% for annual and between 72\% to 64\% for seasonal time series. We include the results of two additional ESMs from CMIP6, the MPI-ESM1-2-HR and the CESM2 model, for comparison with GFDL-ESM4 in the SI (Table S1). The ME of the MPI-ESM1-2-HR model is higher than for GFDL-ESM4 while the CESM2 shows lower bias. The average ME of CEMS2, however, remains higher than our GAN-based post-processed CMCMc-LPJmL model.
        
        In addition to the mean error we also evaluate the difference in the 95th percentile of the precipitation above a threshold of 0.5 [mm/day] per grid cell. The spatial plots are shown in Figs.~S5-S9 and summarized as absolute averages in Table S2. Again, the GAN outperforms the other baseline methods for annual and seasonal time series, reducing biases between 59.76 and 49.11\%.
       
        Also from latitude profiles it can be quantitatively inferred that the GAN outperforms quantile mapping especially regarding the correction of the double-peaked ITCZ, and also that the GAN-processed fields is closer to the ERA5 data than the GFDL-ESM4 simulations, especially in the tropics (Fig.~\ref{fig:temporal_bias}e). 
       
        Regarding the globally averaged temporal distributions, we infer an under-representation of heavy precipitation values in CM2Mc-LPJmL and an over-representation in GFDL-ESM4. 
        QM and our GAN-based method perform similarly well in correcting the distributions over the entire range of precipitation values (Fig.~\ref{fig:temporal_bias}f).

        \begin{table}[h!]
            \centering
            \caption{
                The averaged absolute value of the grid-cell-wise mean error (ME) for the raw CM2Mc-LPJmL and GFDL-ESM4 models, as well as for the QM- and GAN-based post-processing, using the CM2Mc-LPJmL output as input. The bias reduction relative to the raw CMCMc-LPJmL model is given in percentage. Note that the GAN shows the largest reduction of the absolute ME in all cases, with more than 75\% improvement relative to the raw CM2Mc-LPJmL for the annual fields.
                }
            \begin{tabular}{lcccccccc}
               \hline
               Season   & CM2Mc-LPJmL  & GFDL-ESM4 & \%   & QM    & \%   & GAN            & \% \\ 
               \hline
               Annual   & 0.769        & 0.448     & 41.7 & 0.218 & 71.7 & \textbf{0.191} & \textbf{75.2} \\
               DJF      & 0.915        & 0.544     & 40.5 & 0.664 & 27.4 & \textbf{0.256} & \textbf{72}\\
               MAM      & 0.886        & 0.603     & 31.9 & 0.567 & 36.4 & \textbf{0.268} & \textbf{69.8}\\
               JJA      & 0.963        & 0.589     & 38.8 & 0.704 & 26.9 & \textbf{0.270} & \textbf{72} \\
               SON      & 0.823        & 0.508     & 38.3 & 0.552 & 32.9 & \textbf{0.294} & \textbf{64} \\
               \hline
            \label{tab:bias}
            \end{tabular}       
        \end{table} 

    \subsection{Correcting spatial patterns}
    
\begin{figure}[htp!]
            \centering
            \includegraphics[width=1.0\textwidth]{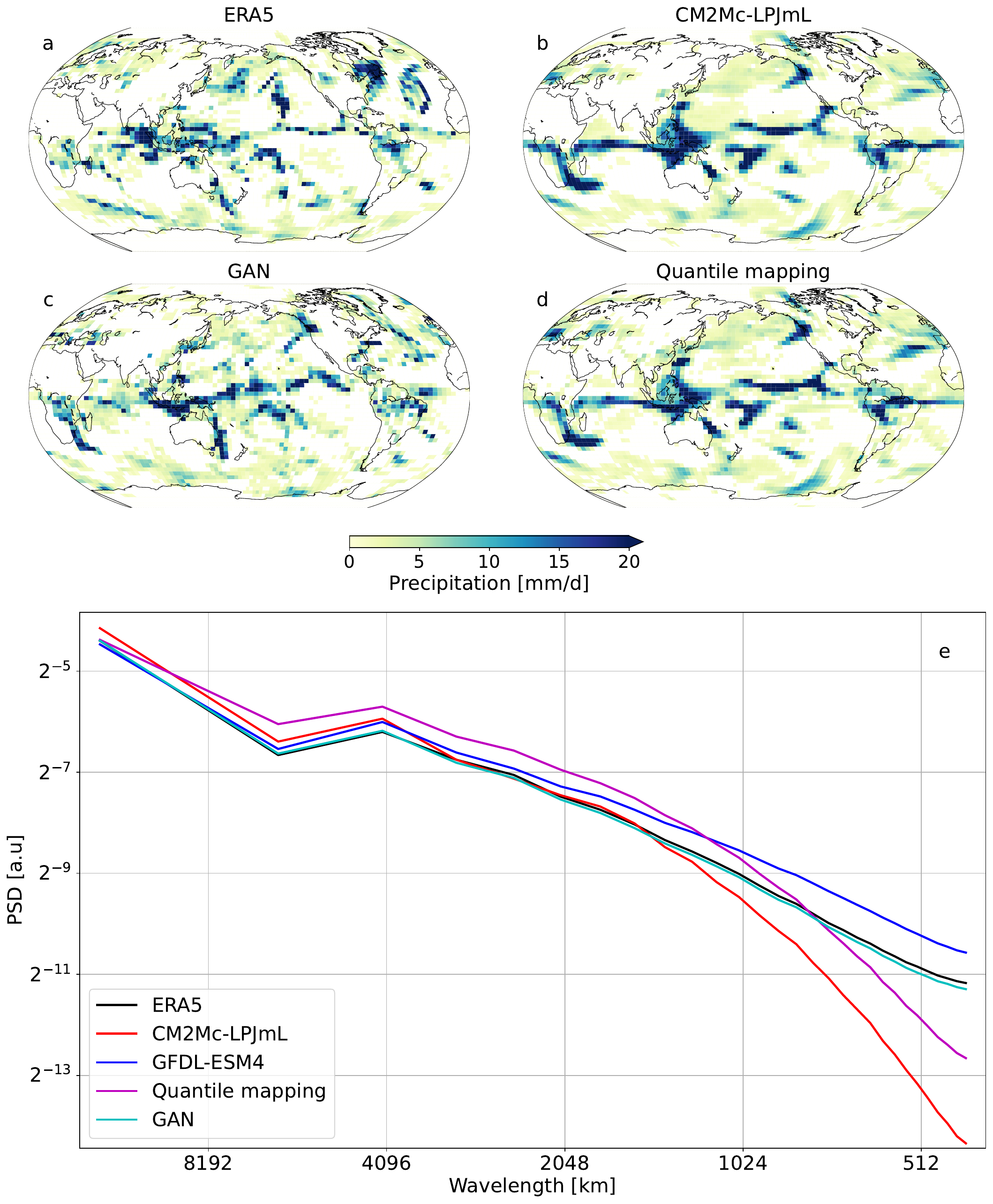}
            \caption{
            Qualitative and quantitaive comparison of the intermittency in daily precipitation above 1~mm/day, on the same date (25th December 2014), for the (a) ERA5 reanalysis, (b) CM2Mc-LPJmL model, (c) GAN-based and (d) QM-based post-processing. The CM2Mc-LPJmL precipitation field (b) corresponds to an input of the GAN-generator which transforms it into the field shown in panel (c). The discriminator network then classifies whether the GAN output (c) or the ERA5 field (a) was generated artificially. Visually, the GAN substantially improves the spatial intermittency seen in ERA5, whereas applying QM does not lead to improved intermittency. Note that the modelled fields are not expected to be point-wise similar to the ERA5 `ground truth' (a), since these are time slices from climate projection runs. 
            (e) The spatial power spectral density (PSD) of the different precipitation fields, averaged radially in space and over time. For ERA5 reanalysis (black), CM2Mc-LPJmL (red), GFDL-ESM4 (blue), quantile mapping (magenta) and the GAN (cyan). Note that only GAN-based post-processing of the CM2Mc-LPJmL model yields an accurate PSD across all spatial scales.}
            \label{fig:spatial_bias}
        \end{figure}

        We continue with assessing the ability of our correction method to improve the spatial structure of the ESM precipitation output. Most importantly, we investigate to which degree the characteristic high-frequency spatial variability of precipitation which is not represented well in the CM2Mc-LPJmL model output, can be improved (see Fig.~\ref{fig:spatial_bias} for some example fields). 
        To quantify this spatial intermittency in the precipitation fields, we compute the radially averaged power spectral density (PSD) following \cite{harris2001multiscale, sinclair2005empirical, ravuri2021skillful}.
        First, the PSD is computed for each daily spatial precipitation field and then the mean is taken over the resulting spectrograms, shown in Fig.~\ref{fig:spatial_bias}e. 
        While the CM2Mc-LPJmL precipitation shows a reduced density at high frequencies (i.e.,  short wavelengths below 1024~km), the GFDL-ESM4 model exhibits an unrealistically high PSD in the same range. Quantile mapping shifts the CM2Mc-LPJmL spectrum towards ERA5, but results in an overshoot in the mid-range and long wavelengths, while the higher frequencies remain underestimated. Only the GAN is able to produce a power spectrum that is consistent with ERA5, especially for short wavelengths, i.e., the high-frequency range that is crucial for precipitation.

    \subsection{Non-stationary climate scenario}
    
        Climate projections under a changing radiative forcing induced by anthropogenic greenhouse gas release constitute an out-of-sample problem: The conditions for which predictions shall be made are different from the conditions for which historical data are available for training. Methods for post-processing or correcting the output of ESMs tasked with such projections hence need to be able to generalize to states that deviate from the historical period, where observations are available.
        Here, we test our GAN approach for the CMIP6 SSP5-8.5 scenario until the end of the 21st century. 
        The SSP5-8.5 ``business as usual'' scenario represents an extreme climate scenario in CMIP6, with the strongest increase in CO$_2$.
        This scenario has been chosen to test how well the GAN model can capture the non-stationarity in this extreme case.
        
\begin{figure}[h!]
            \centering
            \includegraphics[width=0.7\textwidth]{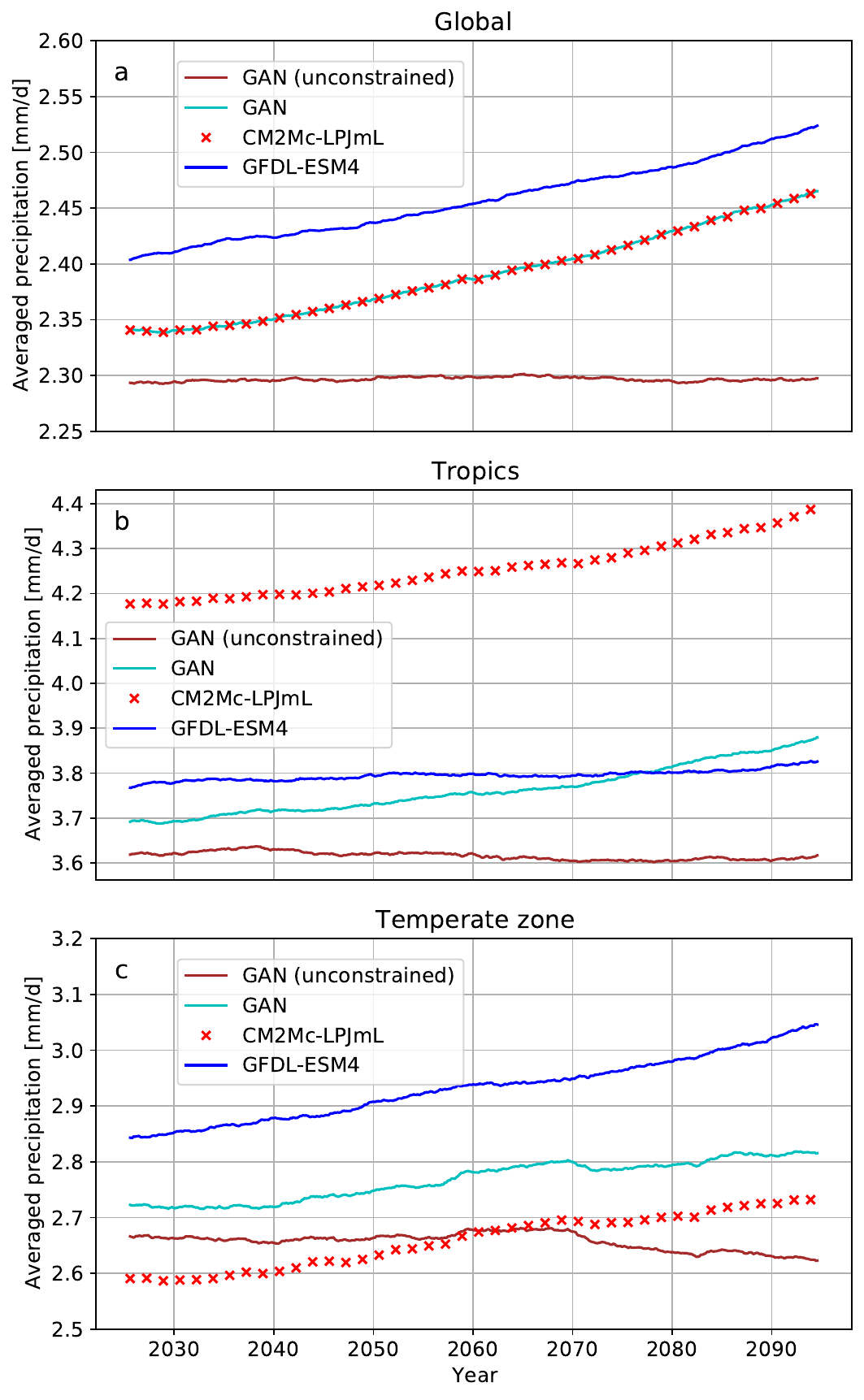}
            \caption{Large-scale trends as a three year rolling-mean of monthly and spatially average precipitation for the CMIP6 SSP5-8.5 scenario. For (a) global data, (b) the tropics and (c) temperate zone, of the CM2Mc-LPJmL (red crosses) and GFLD-ESM4 (blue) models, as well as the constrained (cyan) and unconstrained (brown) GANs. Only by adding the physical constrained to preserve the global precipitation amount per timestep enables the GAN (cyan) to follow the transient dynamics of the non-stationary climate scenario.}
            
            \label{fig:non_stationary}
        \end{figure}

        The CM2Mc-LPJmL and GFDL-ESM4 models both show monotonically increasing global mean precipitation with similar trends over the current century (Fig.~\ref{fig:non_stationary}a), which is in agreement with other studies \cite{ipcc-ar6-wg1-2021}. 
        In contrast, the unconstrained GAN, trained on the historical period, does -- as expected -- not exhibit an increase in average global precipitation, since it is by itself not able to generalize to the changing boundary conditions given by higher greenhouse gas concentrations and temperatures. 
       
        In the tropics ($23\degree$ S to $23\degree$ N), GFDL-ESM4 remains overall lower in mean precipitation than CM2Mc-LPJmL, while also exhibiting a much less pronounced increase over the entire period (Fig.~\ref{fig:non_stationary}b). 
        For the temperate zones from $40\degree$ N/S to $60\degree$ N/S, the GFDL-ESM4 model shows an overall higher mean precipitation with a slightly stronger positive trend than CM2Mc-LPJmL (Fig.~\ref{fig:non_stationary}c). 
        
        By construction of the constraint introduced in Eq.~\ref{eq:constraint}, the GAN-processed precipitation is identical to the increasing global average of the CM2Mc-LPJmL output (Fig.~\ref{fig:non_stationary}a). 
        Without the constraining layer added to the GAN, however, the GAN-processed precipitation stays relatively constant without a substantial trend. 
        In both tropical and temperate zones, the constrained GAN corrects the precipitation towards the more complex and higher-resolution GFDL-ESM4, while following the trend of the CM2Mc-LPJmL model. Again, the unconstrained model remains relatively constant in both cases, with a small decrease over time in the temperate zone. Note that the GFDL-ESM4 does not represent a ground truth, but only one realisation of a possible Earth system trajectory, for comparison. This  can be seen by the differing trends of two other CMIP6 models in Fig.~S13. It should, however, be expected that the precipitation output from the CMIP6 models is much more realistic than the raw precipitation from the comparably low-resolution CM2Mc-LPJmL model. 
        The CMIP6 model GFDL-ESM4 also appears to be calibrated well with respect to large-scale averages over the historical test period, as can be seen in Fig.~S12, in which the GAN shows improvements over the CM2Mc-LPJmL inputs.
        
    \subsection{Interpretability of the GAN-based correction}
        We investigate in the following whether the GAN has learned an ESM output correction that is also physically reasonable.
        The attribution maps are computed with SmoothGrad for each prediction of the discriminator $D_Y$, with daily CM2Mc-LPJmL precipitation fields given as input.
        The discriminator has been trained to distinguish between reanalysis (ERA5) and GAN-processed precipitation fields and we are interested to see which spatial regions in the ESM output the discriminator regards as most important for the distinction.
        These regions then need to be corrected the most by the generator, 
        implying where the most pronounced biases of CM2Mc-LPJmL are.
 
\begin{figure}[htp!]
            \centering
            \includegraphics[width=1.0\textwidth]{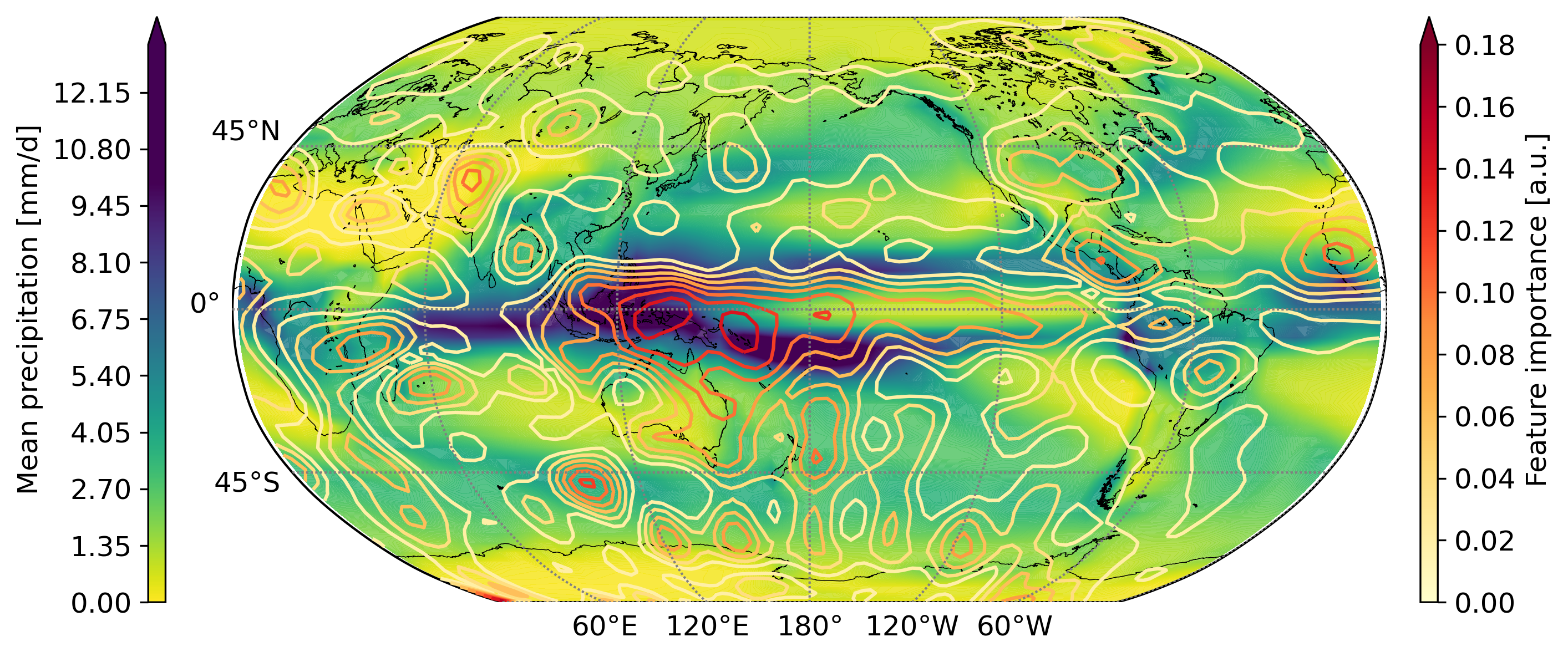}
            \caption{Annual average of daily precipitation fields
            from CM2Mc-LPJmL (color shading with scale according to the colorbar on the left) together with attribution maps (contour lines with color scale according to colorbar on the right). Note that we applied a Gaussian filter to the attribution maps to further reduce the noise.
        A standard deviation $\sigma=1.5$ for the filter was found to give robust results.
        The pacific region in the tropics shows the highest annual mean precipitation, and also the highest feature importance. The same region also exhibits the largest bias of CM2Mc-LPJmL, see in Fig.~\ref{fig:temporal_bias}. 
        Note that especially the double-ITCZ bias is a common and long-standing problem in the precipitation output of many general circulation models \cite{tian2020double}.  
        }
            \label{fig:smooth_grad}
        \end{figure}

        The temporal average of the CM2Mc-LPJmL precipitation is shown in Fig.~\ref{fig:smooth_grad} together with the absolute value of the attribution map as contour lines.
        The regions of highest importance are shown in red and coincide with the region in the western Pacific where the strongest biases and in particular the double-peaked ITCZ of CM2Mc-LPJmL are located (as shown in Fig.~\ref{fig:temporal_bias} and Fig.~S1).
        Although the GAN is trained on daily precipitation fields, it has thus learned to identify regions that show biases occurring on interseasonal to interannual scales. 
    
\section{Discussion}
    
    We have introduced a physically constrained generative adversarial network
    that, combined with the computationally lightweight and efficient CM2Mc-LPJmL Earth system model, is able to produce highly realistic precipitation simulations at low computational costs. 
    
    Our method improves the ESM output in two ways: (i) the temporal distributions of the CM2Mc-LPJmL model precipitation, as well as (ii) the spatial patterns and in particular the spatial intermittency of the CM2Mc-LPJmL model precipitation.
    Our approach is evaluated against quantile mapping \cite{cannon2015bias}
    and the much more advanced CMIP6 GFDL-ESM4 model, \cite{gfdlesm4}
    taking ERA5 reanalysis data as ground truth. Note that any other, and especially purely observational, precipitation dataset with sufficient temporal resolution could readily be used instead.
    
    Given that the training samples are unpaired as a result of the chaotic nature of observed and simulated Earth system trajectories, a comparison of single prediction-target pairs is not possible. We therefore evaluate the GAN performance on long-term summary statistics over the entire test set period.
    When evaluating the skill to improve temporal distributions, we find that our proposed method outperforms both baselines, showing the lowest mean errors and the smallest difference in the 95th precipitation percentile. 
    The improvement over quantile mapping is especially pronounced for seasonal time series, where only our method successfully removes the double-peaked ITCZ of the CM2Mc-LPJmL model. 
    This is in contrast to the results by \cite{franccois2021adjusting}, who report a comparable skill of their CycleGAN implementation with quantile mapping for regional climate simulations.
    Our method corrects relative frequency histograms over the entire range of precipitation values, similarly well to QM, which is designed for this task.
    
    Crucially, our GAN-based approach also improves the spatial structure of the ESM precipitation fields, which is not possible with traditional approaches.
    The GAN yields realistically intermittent spatial patterns that are characteristic for precipitation on all resolved scales, and in this regard outperforms both the quantile-mapping-based post-processing and the comprehensive, high-resolution GFDL-ESM4 model.
    These results show that our method, combined with the computationally lightweight and efficient CM2Mc-LPJmL ESM, can produce precipitation fields that are at least comparable to state-of-the-art, and much more computationally expensive CMIP6 models.
    
    We applied our method to the strongly non-stationary SSP5-8.5 CMIP6 climate scenario until 2100 to test the GAN's ability to capture these non-stationarity and the transient dynamics. The unconstrained GAN trained on observations does not generalize to the unseen climate state. 
    It does not show an increase in global mean precipitation, as one would expect from the thermodynamic Clausius-Clapeyron relation and as seen in the numerical ESMs \cite{allan2008atmospheric, donat2013updated, guerreiro2018detection, traxl2021role}.
    This can be explained by the fact that the precipitation of the future scenario lies well outside the training distribution. 
    To solve this and help the GAN to generalize to this kind of out-of-sample prediction, a physical constraint to preserve the global precipitation amount of the ESM in each time step was introduced as additional network layer in the GAN. 
    The global constraint allows the GAN to improve the precipitation regionally by accounting for local characteristics, while producing the same global mean as the ESM by construction.
    Conserving a physical quantity that is simulated numerically, such as the global precipitation sum in our study, also means that it cannot be improved with respect to observations by definition of the constraint.
    The global precipitation trend can, however, be expected to be represented comparably well in the numerical ESM through thermodynamic processes.
    Adding this constraint enables the GAN to follow the non-stationary, transient dynamics of the SSP5-8.5 scenario. 
    
    The generator architecture in this study is deterministic, producing the same input-output-pairs once it is trained.
    This enables run-to-run reproducibility, where uncertainties of the ESM can then be quantified through ensemble runs. 
    Since the training itself is stochastic, one can create an ensemble to estimate the uncertainties resulting from GAN training (see Fig.~S14).  
    A potential direction for future research could be to develop a stochastic model that directly learns the uncertainties. 
    
    We demonstrate how feature attribution from interpretable Artifical Intelligence can be applied for a GAN, enabling a physical interpretation of this deep learning model.
    We find that the discriminator part of the GAN has learned to identify those regions for its decisions that are critical also from a physical perspective. These regions highlighted by our GAN interpretation  are the ones with the highest absolute errors of the raw CM2Mc-LPJmL, and are known to be the most problematic for ESM precipitation in general.
    Namely, the tropical Pacific Ocean was found to be of highest importance for the discriminator. In this region, the particularly heavy precipitation is often caused by deep convection-driven clouds, which are difficult to model numerically \cite{tian2020double}.
    The sensitivity of the discriminator in the Pacific region
    also explains the effectiveness of our generator network to reduce the double-peaked ITCZ bias. 
    This is the region where the generator needs to modify the CM2Mc-LPJmL precipitation field most in order to avoid rejection by the discriminator.
    The results indicate that the GAN has successfully learned the long-term statistics while being trained on samples of much shorter time scales.
    This makes GANs particularly suitable for climate applications, where training samples and the statistics of interest are often on very different time scales.
    
    The main contribution of our approach is the efficient simulations of highly realistic precipitation fields, by combining a physically constrained GAN with an ESM of reduced complexity. Producing similarly realistic fields purely numerically would require much more computational resources. 
    For comparison, our post-processed CM2Mc-LPJmL ESM takes about 0.5 hours to compute a model year using 28 CPUs, whereas the much more complex GFDL-ESM4 requires 2 hours computational time on 1000 CPUs for a model year \cite{gfdlesm4}. 
    This corresponds to an increased computational efficiency by roughly two orders of magnitude, keeping in mind that GFDL-ESM4 produces higher resolution output.
    The time the GAN post-processing takes is negligible in comparison, taking 0.35 seconds per model year on a V100 GPU and 37.17 seconds on a single CPU. The quantile mapping is similarly efficient taking 0.59 seconds per model year on a CPU.
    
    Based on our findings, there are several directions for extending our method.
    Down-scaling applications that increase the resolution of the ESM could be a direction for future research.
    Conditioning the generator by adding variables that are physically linked to precipitation, such as humidity, temperature, or wind, could further improve our method. 
    The precipitation data, improved by our method, may be used as input to other stand-alone Earth system components such as vegetation, that require realistic climate input.
    
\section*{Acknowledgments}
    The authors would like to thank the referees for their helpful comments and suggestions.
    NB and PH acknowledge funding by the Volkswagen Foundation, as well as the European Regional Development Fund (ERDF), the German Federal Ministry of Education and Research and the Land Brandenburg for supporting this project by providing resources on the high performance computer system at the Potsdam Institute for Climate Impact Research. 
    MD acknowledges funding by the Volkswagen Foundation project POEM-PBSim.
    The authors thank the International Max Planck Research School for Intelligent Systems (IMPRS-IS) for supporting FS.
    NB acknowledges further funding by the Federal Ministry of Education and Research under grant No. 01LS2001A.

\section*{Data availability}
    The ERA5 reanalysis data is available for download at the Copernicus Climate Change Service (C3S) 
    (\url{https://cds.climate.copernicus.eu/cdsapp#!/dataset/reanalysis-era5-single-levels?tab=overview} and \url{https://cds.climate.copernicus.eu/cdsapp#!/dataset/reanalysis-era5-single-levels-preliminary-back-extension?tab=overview}).
    Output data from the CM2Mc-LPJmL model is available at \url{https://doi.org/10.5281/zenodo.4683086} \cite{data-gmd-2020-436}.
    The CMIP6 data can be downloaded at \url{https://esgf-node.llnl.gov/projects/cmip6/}.

\section*{Code availability}
    For the CM2Mc-LPJmL model code see \url{https://doi.org/10.5281/zenodo.4700270} \cite{code-gmd-2020-436}.  The Python code for processing and analysing the data, together with the PyTorch Lightning \cite{falcon2019pytorch, pytorch2019} code for training is available as a compute capsule at Code Ocean: https://doi.org/10.24433/CO.2750913.v1 \cite{Phys_compute_capsule}.

\section*{Competing interests}
    The authors declare no competing interests.

\section*{Authors contribution}
    PH and NB conceived the research and designed the study with input from all authors. PH performed the numerical analysis. MD conducted the CM2Mc-LPJmL experiments. All authors interpreted and discussed the results. PH wrote the manuscript with input from all authors.

\section*{Materials and Methods}
    
    \subsection*{The Earth system model CM2Mc-LPJmL}

    The coupled Earth system model CM2Mc-LPJmL v1.0 \cite{druke2021cm2mc} combines the coarse-grained but relatively fast atmosphere and ocean model CM2Mc \cite{Galbraith2011ClimateModel} with the state-of-the-art dynamic global vegetation model (DGVM) LPJmL5 \cite{Schaphoff2018LPJmL4Description, Schaphoff2018LPJmL4Evaluation, VonBloh2018Implementing5.0}.

    CM2Mc is a coarser (3°x3.75° latitude-longitude) configuration of the Climate Model CM2 \cite{Milly2002GlobalModel}, which has been developed at the Geophysical Fluid Dynamics Laboratory (GFDL). The original configuration of CM2Mc includes the Modular Ocean Model 5 (MOM5) and the global atmosphere and land models AM2-LM2 or AM2-LM \cite{Anderson2004TheSimulations} with static vegetation. In CM2Mc-LPJmL, the land component LM/LM2 is replaced by the dynamic global vegetation model LPJmL5, while AM2 and MOM5 remain dynamically coupled to the model framework. The Flexible Modeling System (FMS) developed by GFDL connects all different model compartments and calculates the fluxes between them.\\
    The state-of-the-art and thoroughly validated DGVM LPJmL (Lund-Potsdam-Jena managed Land) simulates global surface energy balance, water fluxes and carbon stocks and fluxes for natural and managed land. Being forced by climate and soil data, LPJmL simulates the impact of bioclimatic limits and effects of heat, productivity and fire on plant mortality to determine the establishment, growth, competition and mortality for different plant functional types (PFTs) in natural vegetation and crop functional types (CFTs) on managed land. Since its original implementation \cite{Sitch2003EvaluationModel} the model now incorporates a water balance \cite{Gerten2004Jan}, agriculture \cite{Bondeau2007Mar}, wildfire in natural vegetation \cite{Thonicke2010Jun,Druke2019ImprovingData}, and the impact of multiple climate drivers on phenology \cite{Forkel2014IdentifyingIntegration,Forkel2019ConstrainingObservations}.

    In CM2Mc-LPJmL, the fluxes simulated by LPJmL depend, of course, on the precipitation modelled by AM2. As a stand-alone model LPJmL has been mainly calibrated with respect to reanalysis, and a similarly accurate precipitation output within CM2Mc-LPJmL would hence be favorable to maintain consistency and to obtain realistic surface fluxes from LPJmL. For the overall performance of CM2Mc-LPJmL, realistically simulated precipitation fields are therefore crucial. This motivates the work presented below, where we use a specific kind of GAN to transform the AM2 precipitation fields toward fields that are indistinguishable from ERA5 precipitation fields (see below).
    
    The model experiments of this paper are consistent with \cite{druke2021cm2mc}. After a 5000-year stand-alone LPJmL spin-up, a second fully coupled spin-up under pre-industrial conditions without land use was performed for 1250 model years. In this way we ensure that the model starts from a consistent equilibrium between the long-term soil carbon pool, vegetation, ocean, and climate.
    
    The subsequent transient historic phase of the model is performed from 1700-2018, using historic land use data from 1700 \cite{Fader2010Apr} and historic concentrations of greenhouse gases, solar radiation, ozone concentrations and aerosols from 1860, which were kept at pre-industrial conditions beforehand.
    
    From 2019 until 2100 the model is forced by constant land use from the year 2018 and CO$_2$ equivalents of the atmospheric forcing prescribed in the CMIP6 SSP5-8.5 (``business as usual'') climate scenario that assumes a continued increase in CO$_2$ emissions.
    
    \subsection*{Cycle-consistent generative adversarial networks}
         
        Generative adversarial networks (GANs) are designed to learn a target distribution $p_y(y)$ through a two-player ``minimax'' game between a generator $G$ and a discriminator $D$ \cite{goodfellow2014generative}.
        The generator network is trained to transform an input $x \in X$ to values that approximate samples from a target domain $y \in Y$, i.e. the generator is trained to learn the mapping $G: X \rightarrow Y$.
        Samples from the generator and the target dataset are then shown to the discriminator, which classifies their origin.
        In this way, the generator and discriminator compete against each other, thereby improving the quality of the generated samples. 
        The training can be formulated as 
        
        \begin{equation}
            G^* = \underset{G}{\mathrm{min}} \; \underset{D}{\mathrm{max}} \; \mathcal{L}_{GAN}(D,G),
        \end{equation}
        where $G^*$ is the optimal generator and $\mathcal{L}_\mathrm{GAN}(D,G)$ is the loss function defined as
        
        \begin{equation}
            \mathcal{L}_\mathrm{GAN}(D,G) = \mathbb{E}_{y\sim p_y(y)}[\log ( D(y) )] + \mathbb{E}_{x\sim p_{x}(x)}[\log ( 1- D(G(x)))].
        \end{equation}
        In our situation, $X$ and $Y$ correspond to the sets containing precipitation fields from the CM2Mc-LPJmL Earth system model and ERA5 reanalysis, respectively (samples are shown in Fig.~\ref{fig:spatial_bias}). 
        In the above formulation, GANs have often been found to suffer from instabilities and difficulties to generalize to distributions of higher dimensionality, such as in image-to-image translation without pairwise matching samples.
        One reason for the instabilities is the highly under-constrained mapping to be learned by the generator. 
        To alleviate this problem, cycle-consistent GANs have been proposed recently \cite{zhu2017unpaired}.
        They aim to constrain the space of mappings by training a second pair of generator and discriminator networks, which learns the inverse mapping  $F: Y \rightarrow X$.
        A schematic of the cycle-consistent GAN model is shown in Fig.~\ref{fig:gan_setup}.
        Both generators should perform bijective (i.e., one-to-one) mappings \cite{zhu2017unpaired} and are therefore trained at the same time, together with a regularization term that enforces consistency of translation cycles, i.e. $x \rightarrow G(x) \rightarrow F(G(x)) \approx x$ and vice versa for $y$.
        The corresponding loss functions are then
        
        \begin{align}
            \mathcal{L}_{X \rightarrow Y}(G, D_Y) & = \mathbb{E}_{y\sim p_{y}(y)}[\log( D_Y(y) )] \\
            & + \mathbb{E}_{x\sim p_{x}(x)}[\log(1 - D_Y(G(x)))], \nonumber
        \end{align}
        and similarly, 
        
        \begin{align}
            \mathcal{L}_{Y \rightarrow X}(F, D_X) & = \mathbb{E}_{x\sim p_{x}(x)}[ \log( D_X(x))] \\
            & + \mathbb{E}_{y\sim p_{y}(y)}[\log(1 - D_X(F(y)))]. \nonumber
            \label{eq:cycle_loss}
        \end{align}
        The cycle-consistency loss is given by
        
        \begin{align}
           \mathcal{L}_{cycle}(G, F) & = \mathbb{E}_{x \sim p_{x}(x)}[|| F(G(x)) - x||_1] \\
           & + \mathbb{E}_{y\sim p_{y}(y)}[|| G(F(y)) - y||_1]. \nonumber
        \end{align}
        The full loss function then reads
        
        \begin{align}
           \mathcal{L}(G, F, D_X, D_Y) = & \mathcal{L}_{X \rightarrow Y}(G,D_Y)  \nonumber\\
           + & \mathcal{L}_{Y \rightarrow X}(F, D_X) \\ 
           + & \lambda \mathcal{L}_{cycle}(G,F), \nonumber
        \end{align}
        which is solved as
        
        \begin{align}
          G^*, F^* = \underset{G,F}{\mathrm{min}} \; \underset{D_X, D_Y}{\mathrm{max}} \;  \mathcal{L}(G, F, D_X, D_Y).
        \end{align}
        We adopt the architecture from \citeA{zhu2017unpaired} and optimize the networks with Adam \cite{kingma2014adam}, using a learning rate of $2e^{-4}$ for both the generator and the discriminator networks and set $\lambda = 10$. 
        Following \citeA{zhu2017unpaired} we set the batch size to 1 and
        train the models for 250 epochs, logging the 50 best performing generators every 10 epochs. 
        The training takes about 5.25 days on a NVIDIA V100 GPU with 32\,GB memory.
        After training the final generator is determined by evaluation on the test set.

    \subsection*{Neural network architectures}
       
        The generator architecture is based on a variant of convolutional residual networks \cite{he2016identity}.
        Convolutional neural networks (CNNs) are commonly employed to process image data. 
        CNNs transform the input data through stacked layers of trainable convolutional filters that are followed by a non-linear activation functions  thereby learning to extract spatial patterns. For a more detailed introduction see, e.g., \cite{goodfellow2016deep}.
        Adopting the naming convention from \cite{johnson2016perceptual, zhu2017unpaired}.
        \textsf{c7s1-k} denotes a layer with a $7\times 7$ convolution followed by instance normalization and ReLU activation with $k$ filters, a stride 1 and reflection padding. 
        \textsf{dk} represents a layer with $3\times 3$ convolutions, instance normalization, ReLU activation, $k$ filters and stride 2.
        \textsf{Rk} are residual blocks with a $3 \times 3$ convolutional layer and $k$ filters.
        \textsf{uk} denots a layer with $3 \times 3$ fractional-strided convolutions, instance normalization, ReLU activation, $k$ filters and stride $1/2$.
        The generator architecture with 6 residual blocks is then
        
        \begin{equation*}
         x_{\mathrm{in}} \rightarrow \textsf{c7s1-64} \rightarrow \textsf{d128} \rightarrow \textsf{d256} \rightarrow \underbrace{[\textsf{R256} \rightarrow]}_{\times 6} \textsf{u128} \rightarrow\textsf{u64} \rightarrow \textsf{c7s1-3} \rightarrow y_{\mathrm{out}},
        \end{equation*}
        where $x_{\mathrm{in}}$ is the input of the generator and $y_{\mathrm{out}}$ the output.
        The discriminator architecture is based on the PatchGAN \cite{Isola2017}. Denoting a $4 \times 4$ convolutional layer with $k$ filters, instance normalization (except for the first layer), leaky ReLU with slope $0.2$ and a stride of $2$ with \textsf{Ck}. 
        The full architecture of the discriminator is
        
        \begin{equation*}
         x_{\mathrm{in}} \rightarrow \textsf{C64} \rightarrow \textsf{C128} \rightarrow \textsf{C256} \rightarrow \textsf{C512} \rightarrow y_{\mathrm{out}}.
        \end{equation*}
        
    \subsection*{Generator constraint}
    
        To enable a better generalization of the GAN to climate states not seen during training, and hence in particular to address the out-of-sample problem imposed by the changing radiative forcing due to anthropogenic greenhouse gas emissions, we introduce the physical constraint of preserving the total global precipitation amount of the CM2Mc-LPJmL model input.
        That is, we add an additional layer to the generator network after training, which re-scales each output $y_i$ at each grid point $i$ as
        
        \begin{equation} 
            \label{eq:constraint}
            \tilde{y}_{i} = y_{i} \frac{\sum_i^{N_{\textrm{grid}} } x_{i}}{\sum_i^{N_{\textrm{grid}} }  y_{i}},
        \end{equation}
        where $N_{\textrm{grid}}$ is the total number of grid-points, $x_{i}$ the CM2Mc-LPJmL precipitation input and $\tilde{y}_i$ the constrained output.
        The motivation of the constraint is that it gives the GAN freedom to change the precipitation locally and to redistribute it in space, while forcing it to follow the global trend prescribed by the ESM. The global trend has been found to be well represented in the ESM, where noise and and biases found on small time and spatial scales are averaged out \cite{druke2021cm2mc}. Also in observations, it has recently been shown that the physically based Clausius-Clapeyron relation, suggesting a 7\% increase in precipitation per degree of warming, holds very well in terms of global averages, despite pronounced regional deviations \cite{traxl2021role}.
        
    \subsection*{Training}
    
        We use daily precipitation from the European Center for Medium-Range Weather Forecasts (ECMWF) Reanalysis v5 (ERA5) product \cite{hersbach2020era5} as a training target and ground truth for evaluation.
        This reanalysis is produced by the Copernicus Climate Change Service (C3S) at ECMWF, combining a large range of satellite- and land-based observations with high-resolution simulations through state-of-the-art data assimilation techniques \cite{courtier1994strategy, hersbach2020era5}.
        The original resolution is 30km horizontally in space and hourly in time, spanning the period from 1950 to present.
        For this study the data is aggregated to daily precipitation sums and re-gridded, following \cite{rasp2020weatherbench, beck2019mswep}, by  bilinear interpolation using the xESMF package \cite{jiawei_zhuang_2020_3700105}, in order to match the resolution of CM2Mc-LPJmL. 
        We split the ESM and ERA5 datasets into the training period 1950-2000 and the test period 2001-2014 (for which also the GFDL-ESM4 data is available), with 18615 and 5110 daily samples, respectively.
        Model simulations from 2019-2100 are used to test the generalization of the network with a CO$_2$ forcing according the CMIP6 SSP5-8.5 (``business as usual'') climate scenario, which assumes a continued increase in CO$_2$ emissions.
        Following \citeA{zhu2017unpaired}, we replace the log likelihood by a least-squares loss, which has been found to improve the training.
        The GAN loss in Eq.~2 is then minimized by both $G$ and $D$, with a loss $\mathbb{E}_{x\sim p_{x}(x)}[( D(G(x))-1)^2]$ for $G$ and
        $\mathbb{E}_{y\sim p_{y}(y)}[( D(y)-1)^2] + \mathbb{E}_{x\sim p_{x}(x)}[( D(G(x)))^2]$ for the discriminator $D$.
        We apply a log-transform to the input data with $\tilde{x} = \log(x + \epsilon) - \log(\epsilon)$ following \cite{rasp2021data}, where $\tilde{x}$ is the transformed precipitation and $\epsilon = 0.0001$. We further normalize the data to the interval $[-1, 1]$, which was found to improve the training performance.
        Once trained, the generator takes only about ten seconds on a NVIDIA V100 GPU to process the test set ESM precipitation.

    \subsection*{Baselines}
    
        We compare our method to quantile mapping, implemented with the xClim package \cite{logan_travis_2021_5649661}, and also carry out comparisons to the raw output of the more advanced CMIP6 climate model GFDL-ESM4 \cite{gfdlesm4}.
        The latter uses AM4 \cite{zhao2018gfdl, zhao2018gfdl2}, a more recent and substantially more complex version of the atmosphere model AM2 used in CM2Mc-LPJmL \cite{gfdl2004new}, with a substantially higher spatial resolution and strongly improved parameterizations of subgrid-scale processes. These improvements of course come at the expense of substantially increased computational costs. The motivation here is to see whether a comparably simple atmospheric general circulation model (GCM) such as AM2 can be combined with the proposed GAN model in order to yield similar results as a comprehensive state-of-the-art atmospheric GCM such as AM4, at a fraction of the computational costs.\\
        Quantile mapping uses the empirical cumulative distribution functions of simulated and observed precipitation to transform the simulated values into the corresponding quantiles derived from observations.
        Before computing the cumulative distribution function, following \cite{cannon2015bias}, we detrend the historical time series, assuming a linear trend.\\
        As an error metric to compare our methods we apply the mean error (ME), which is defined as 
        \begin{equation}
            \textrm{ME} = \frac{1}{N} \sum_{t=1}^{N_{time}} (x_t-y_t) = \frac{1}{N} \sum_{t=1}^{N_{time}}x_t - \frac{1}{N} \sum_{t=1}^{N_{time}}y_t,
            \label{eq:me_bias}
        \end{equation}
        where $x_t$ and $y_t$ are the simulated and observed precipitation at time $t$ for a given grid cell and $N_{time}$ the number of time steps in the test set. Note that the ME is used to evaluate the differences in the time averages per grid cell, as can be seen on the right-hand side of Eq.~\ref{eq:me_bias}.

    \subsection*{Model transparency}
    
        Neural network models are often regarded as black boxes.
        Since it is important for many applications to be able to explain the neural network's prediction, the emergent fields of interpretable \cite{murdoch2019interpretable, toms2020physically} and explainable Artificial Intelligence \cite{sundararajan2017axiomatic, montavon2019layer} aim to improve the transparency.

        Many methods for interpreting neural networks are specifically designed for classification problems \cite{goodfellow2016deep}.
        In the GAN framework, the discriminator network performs such a classification task in distinguishing between generated and real images. 
        Hence, suitable interpretability methods can be applied, even though entire GAN is build for the much more complex generative task.
        Being able to interpret the GAN increases the transparency and trust, since it ensures that the model has learned to identify physically reasonable input features.
        To our knowledge, we are the first to apply an interpretability method in such a way, i.e., to test the physical consistency of the GAN training.
        
        Here, we use the gradient-based method SmoothGrad \cite{smilkov2017smoothgrad} to interpret the discriminator network $D_Y$ that has learned to classify ERA5 and generated precipitation fields. 
        An attribution map $\phi$ is computed by taking the gradient of the neural network $D_Y$ with respect to its input $y$,
        \begin{equation}
            \phi(D_Y, y) = \frac{\partial D_Y(y)}{\partial y}, 
        \end{equation}       
        showing for each input grid cell how much the prediction will change with respect to its input, i.e. how sensitive it is to perturbations of the input.
        It has been observed that using only the gradient of the input, however, tends to give rather noisy attribution maps. 
        Therefore, \citeA{smilkov2017smoothgrad} proposed a technique to reduce the noise, by adding it to the network's input and averaging the gradient over a sample size, e.g. here $N=10$, as
        
        \begin{equation}
            \hat{\phi}(D_Y, y) = \frac{1}{N}\sum^N_{i=1} \phi \left(y+\epsilon_i \right),
        \end{equation}       
        where the noise is sampled from a Gaussian distribution $\epsilon_i \sim \mathcal{N} \left(0,\sigma^2 \right)$.


%
%
\bibliography{bibliography,references_markus}

\end{document}


%
%



\title{\center{Supporting Information for "Physically Constrained Generative Adversarial Networks for Improving Precipitation Fields from Earth System Models"}}
%
%

%
%


\vspace{1cm}
\authors{ Philipp Hess\affil{1,2}, Markus Drüke\affil{2}, Stefan Petri\affil{2}, Felix M. Strnad\affil{2,3}, and Niklas Boers\affil{1,2,4}}

\affiliation{1}{\small Technical University Munich, Munich, Germany; School of Engineering \& Design, Earth System Modelling}
\affiliation{2}{Potsdam Institute for Climate Impact Research, Member of the Leibniz Association, Potsdam, Germany}
\affiliation{3}{Cluster of Excellence - Machine Learning for Science, Eberhard Karls Universität Tübingen, Germany}
\affiliation{4}{Global Systems Institute and Department of Mathematics, University of Exeter, Exeter, UK}

%
%

%

\begin{article}

%
%

\noindent\textbf{Contents}
\begin{enumerate}
\item Figures S1-S14 
\item Table S1-S2
\end{enumerate}


\clearpage

\end{article}


%
%
%

\begin{figure}[ht]
    \centering
    \includegraphics[width=\textwidth]{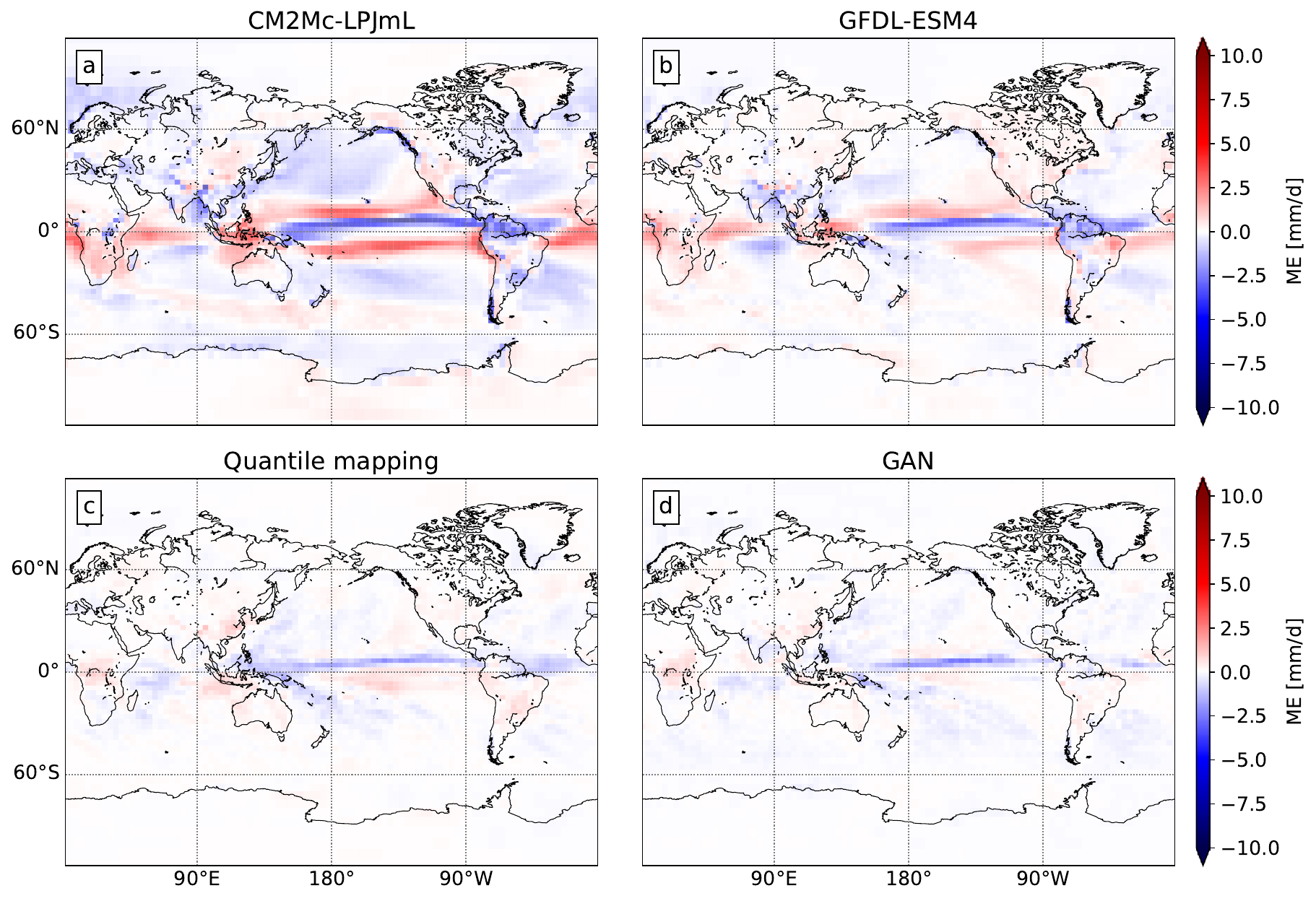}
    \caption{Global maps showing the mean error for the entire test set. For (a) CM2Mc-LPJmL, (b) GFDL-ESM4, (c) QM-based and (d) GAN-based post-processing methods applied to the CM2Mc-LPJmL output.}
\label{fig:annual_bias}
\end{figure} 
        

\begin{figure}[ht]
    \centering
    \includegraphics[width=\textwidth]{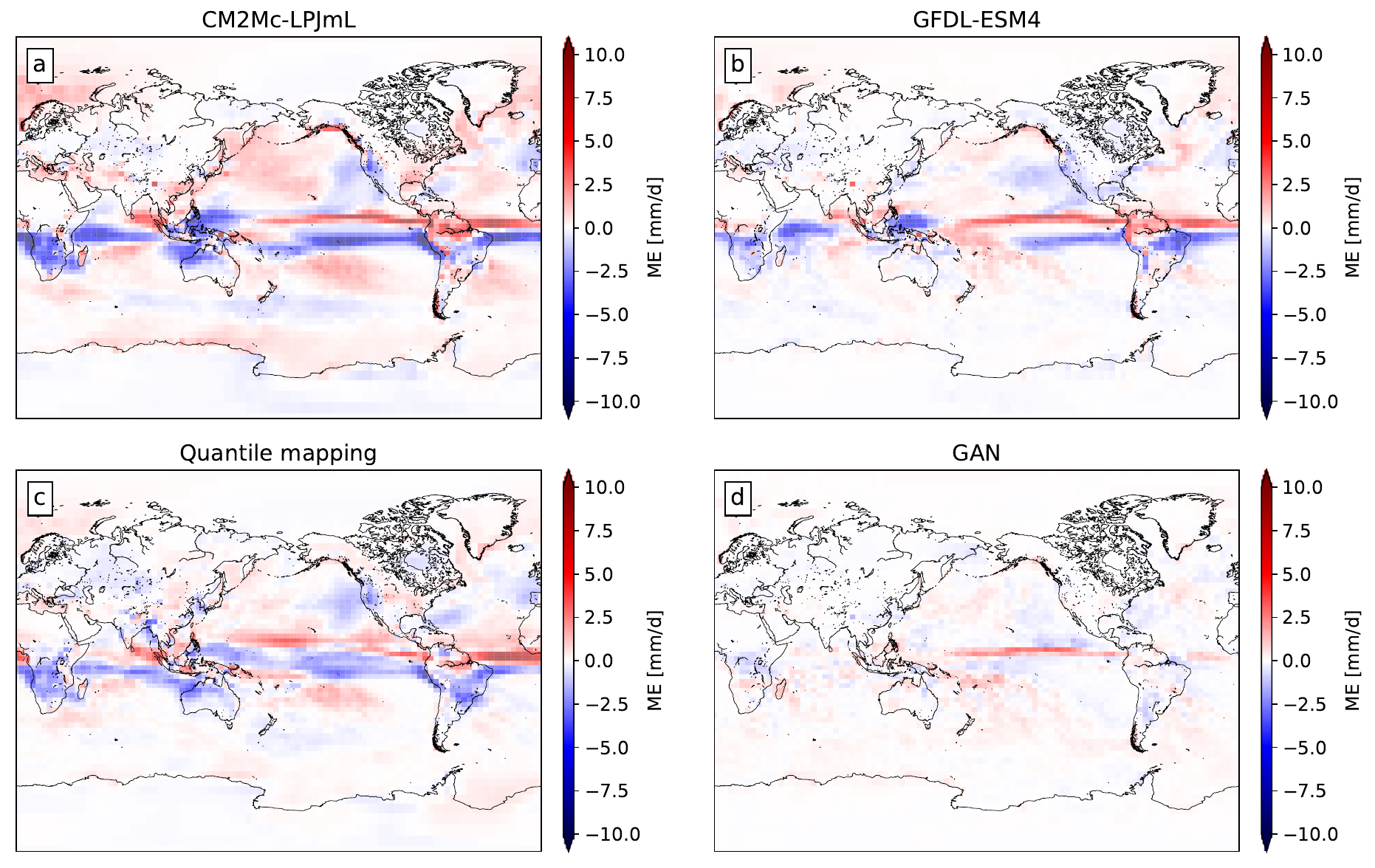}
    \caption{Global maps showing the mean error for the DJF season of the test set. For (a) CM2Mc-LPJmL, (b) GFDL-ESM4, (c) QM-based and (d) GAN-based post-processing methods applied to the CM2Mc-LPJmL output.}
\label{fig:dfj_bias}
\end{figure} 


\begin{figure}[ht]
    \centering
    \includegraphics[width=\textwidth]{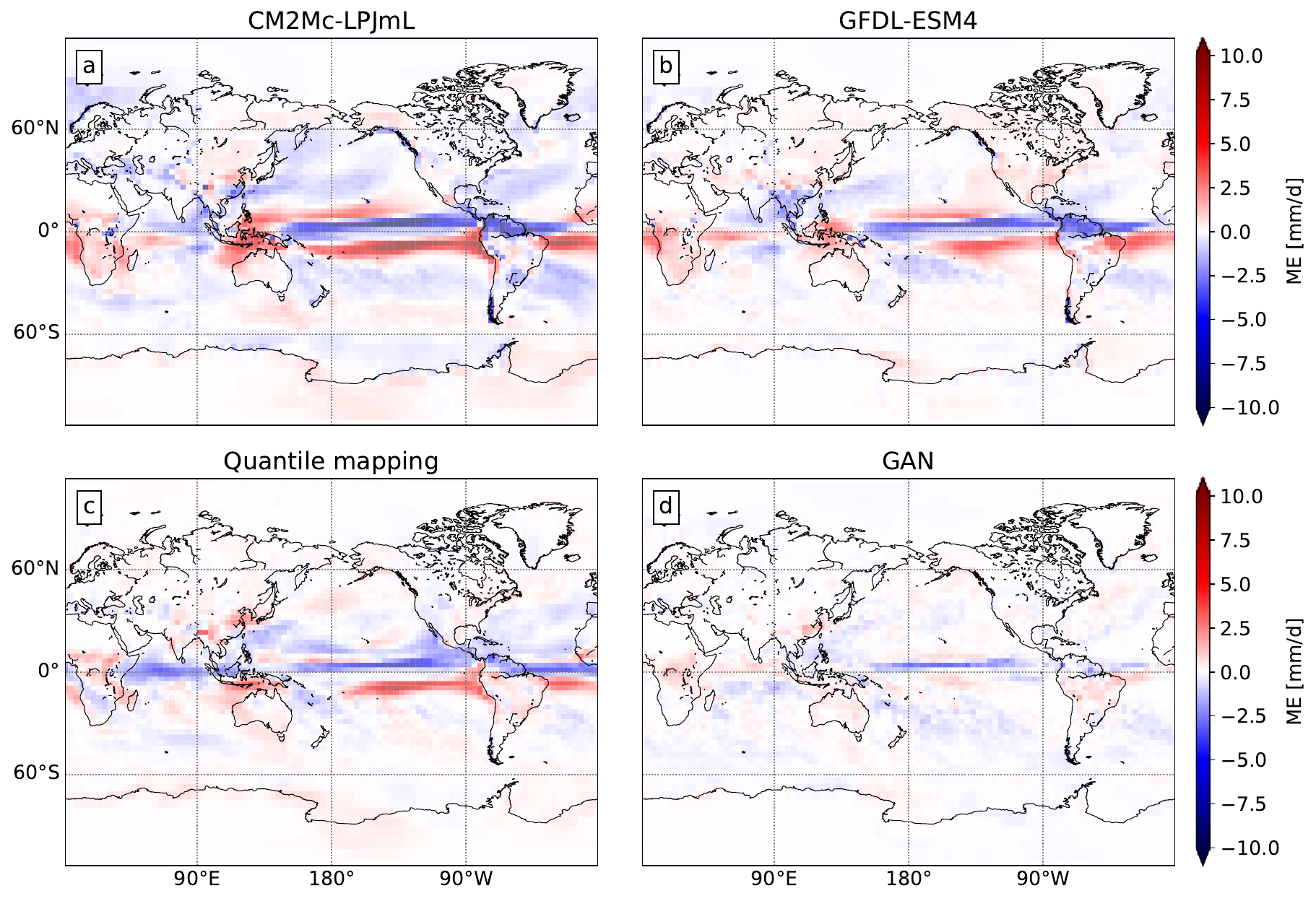}
    \caption{Global maps showing the mean error for the MAM season of the test set. For (a) CM2Mc-LPJmL, (b) GFDL-ESM4, (c) QM-based and (d) GAN-based post-processing methods applied to the CM2Mc-LPJmL output.}
\label{fig:mam_bias}
\end{figure} 


\begin{figure}[ht]
    \centering
    \includegraphics[width=\textwidth]{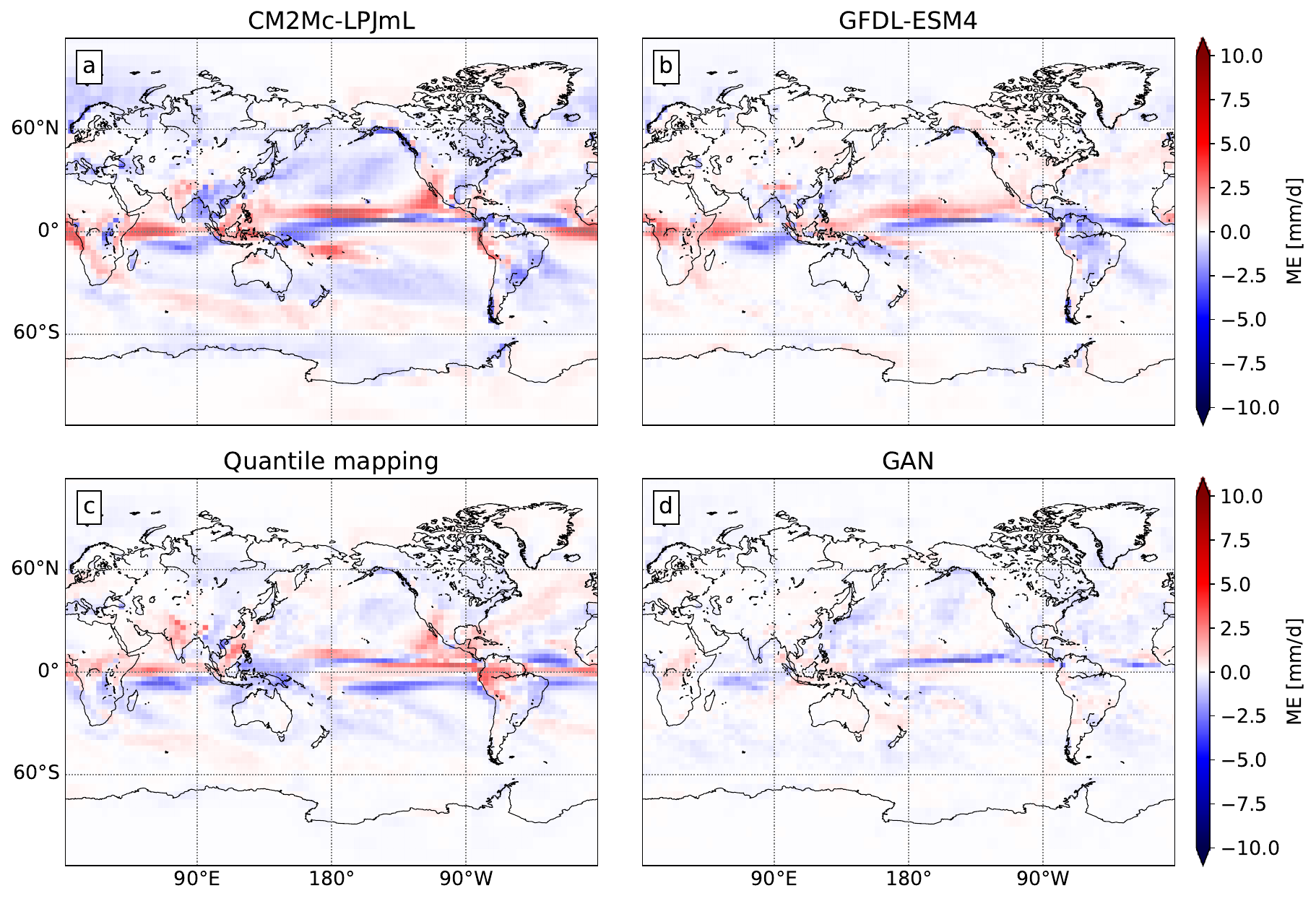}
    \caption{Global maps showing the mean error for the SON season of the test set. For (a) CM2Mc-LPJmL, (b) GFDL-ESM4, (c) QM-based and (d) GAN-based post-processing methods applied to the CM2Mc-LPJmL output.}
\label{fig:son_bias}
\end{figure} 

\begin{figure}[ht]
    \centering
    \includegraphics[width=\textwidth]{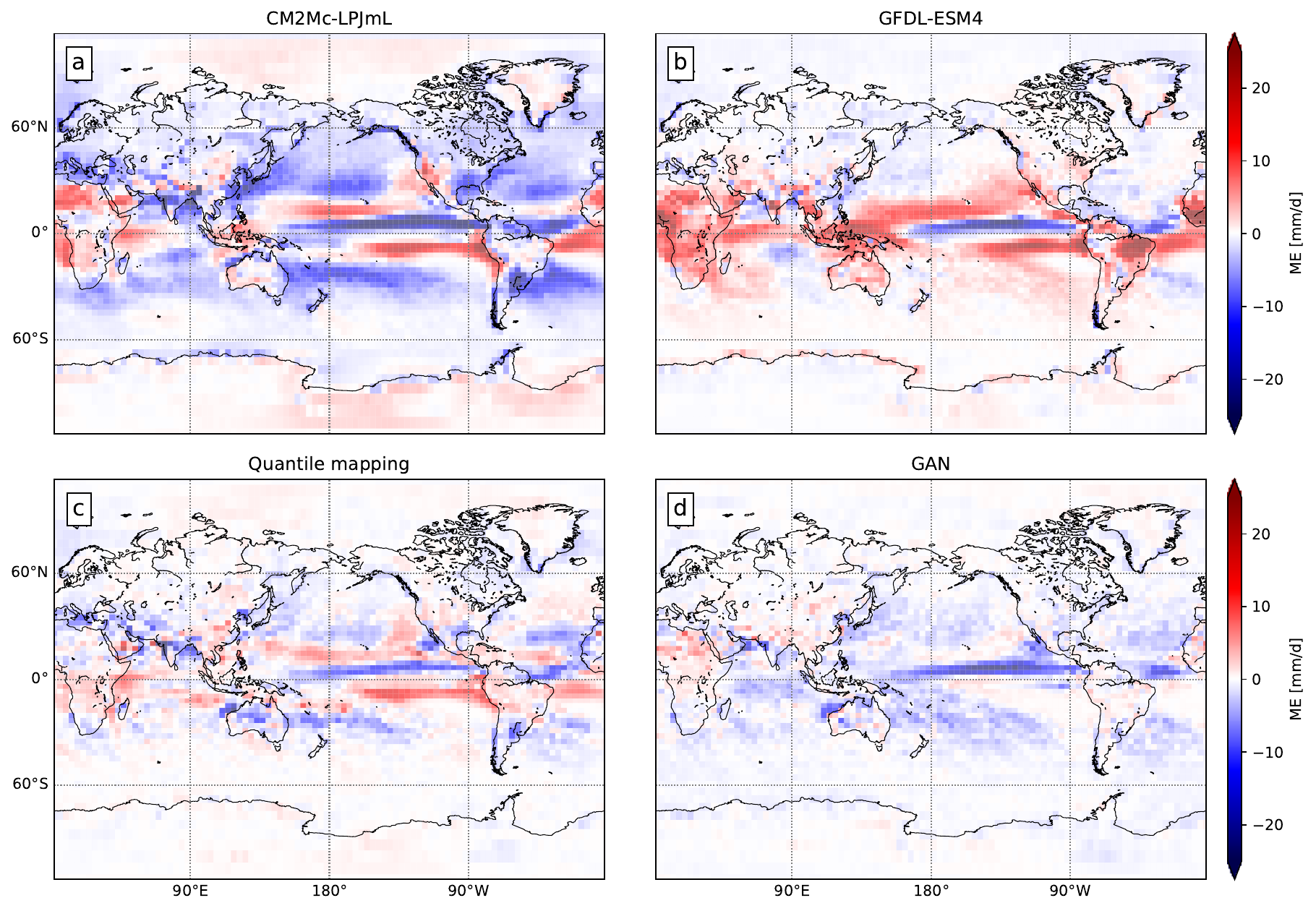}
    \caption{Global maps showing the difference in the 95th precipitation percentile for the annual time series of the test set. For (a) CM2Mc-LPJmL, (b) GFDL-ESM4, (c) QM-based and (d) GAN-based post-processing methods applied to the CM2Mc-LPJmL output. Grid cells where the percentiles could not be determined due to insufficient statistics are shown in grey.}
\label{fig:annual_extremes_bias}
\end{figure} 

\begin{figure}[ht]
    \centering
    \includegraphics[width=\textwidth]{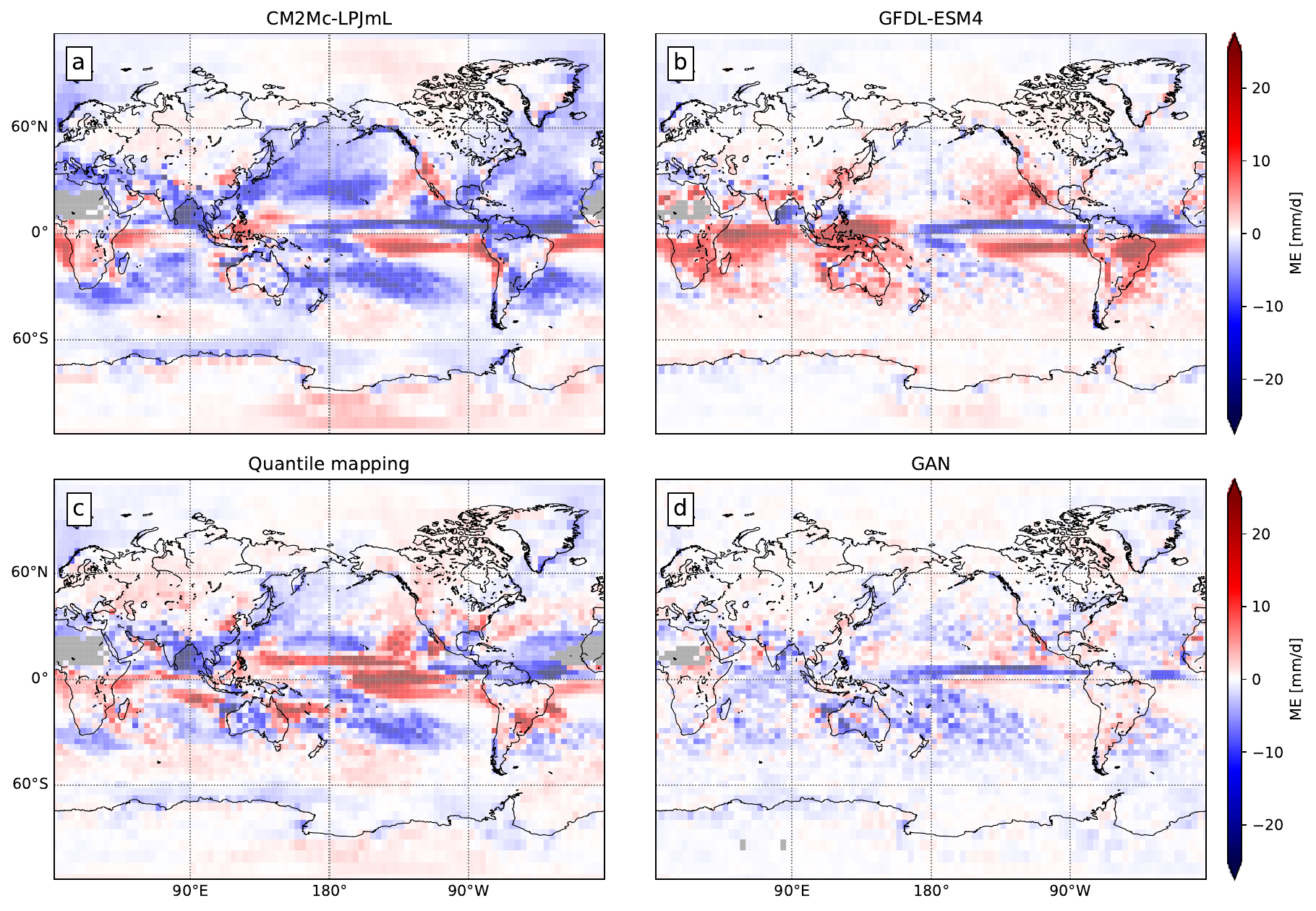}
    \caption{Global maps showing the difference in the 95th precipitation percentile for the DJF season of the test set. For (a) CM2Mc-LPJmL, (b) GFDL-ESM4, (c) QM-based and (d) GAN-based post-processing methods applied to the CM2Mc-LPJmL output. Grid cells where the percentiles could not be determined due to insufficient statistics are shown in grey.}
\label{fig:djf_extremes_bias}
\end{figure} 

\begin{figure}[ht]
    \centering
    \includegraphics[width=\textwidth]{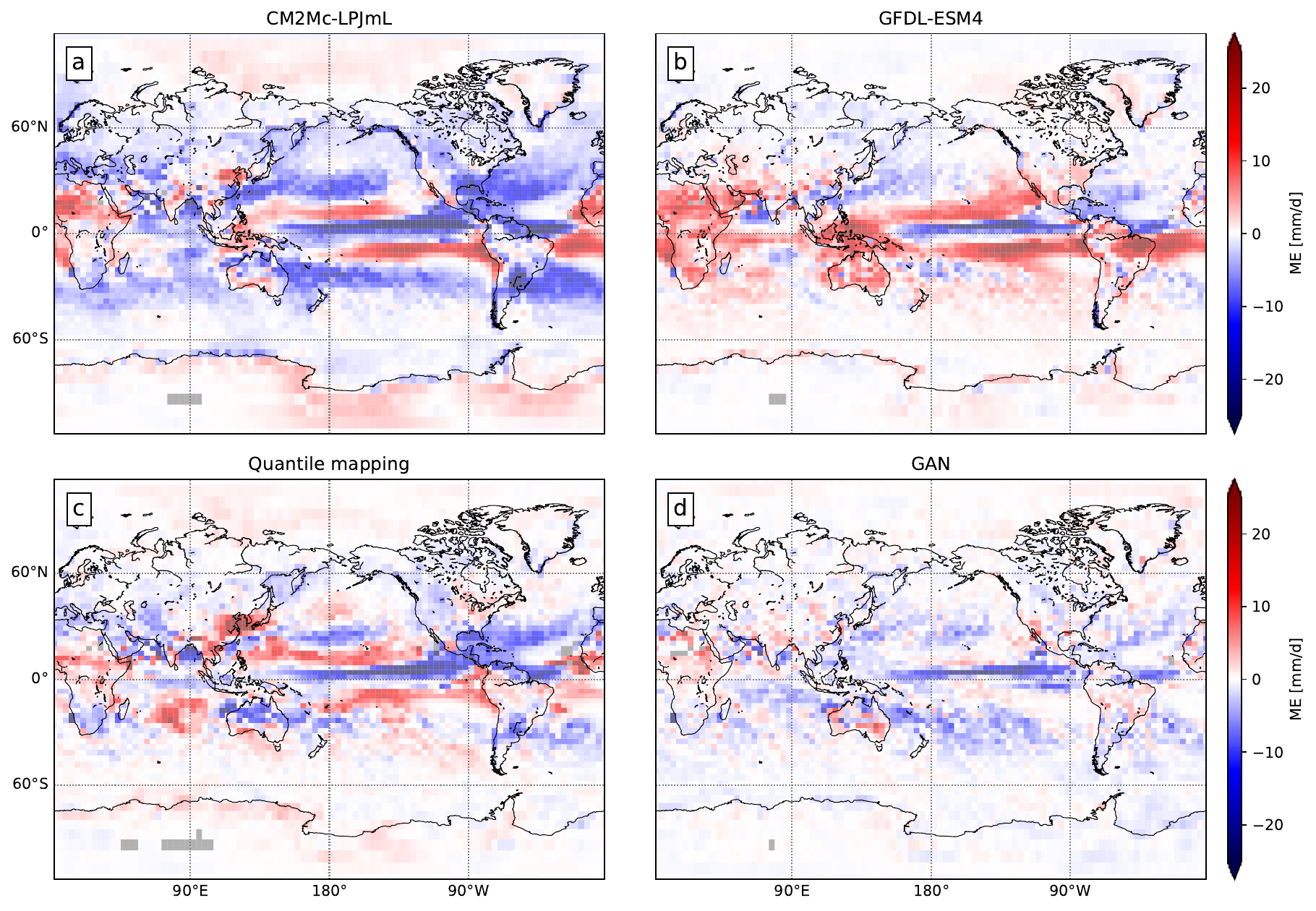}
    \caption{Global maps showing the difference in the 95th precipitation percentile for the MAM season of the test set. For (a) CM2Mc-LPJmL, (b) GFDL-ESM4, (c) QM-based and (d) GAN-based post-processing methods applied to the CM2Mc-LPJmL output. Grid cells where the percentiles could not be determined due to insufficient statistics are shown in grey.}
\label{fig:mam_extremes_bias}
\end{figure} 

\begin{figure}[ht]
    \centering
    \includegraphics[width=\textwidth]{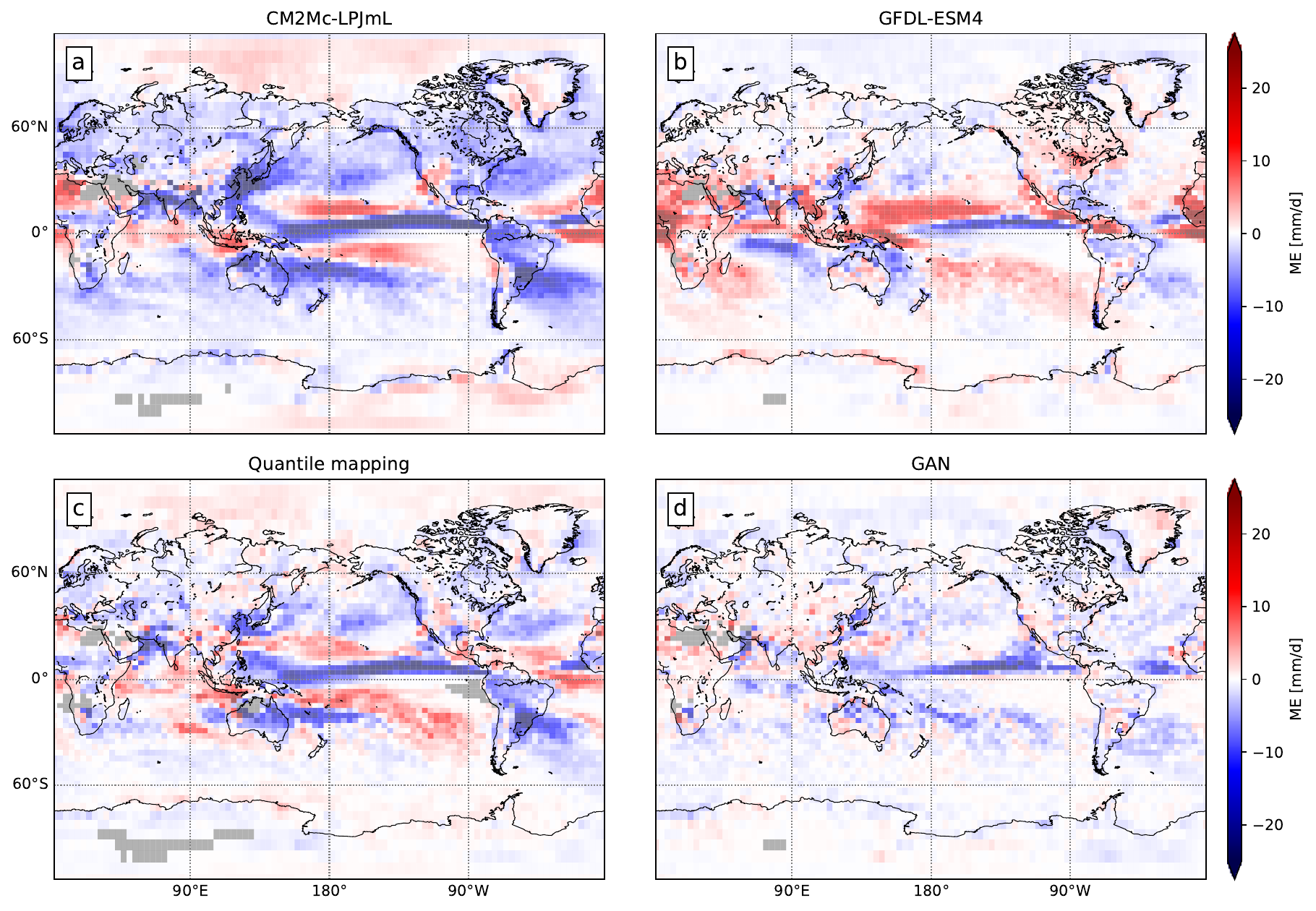}
    \caption{Global maps showing the difference in the 95th precipitation percentile for the JJA season of the test set. For (a) CM2Mc-LPJmL, (b) GFDL-ESM4, (c) QM-based and (d) GAN-based post-processing methods applied to the CM2Mc-LPJmL output. Grid cells where the percentiles could not be determined due to insufficient statistics are shown in grey.}
\label{fig:jja_extremes_bias}
\end{figure} 

\begin{figure}[ht]
    \centering
    \includegraphics[width=\textwidth]{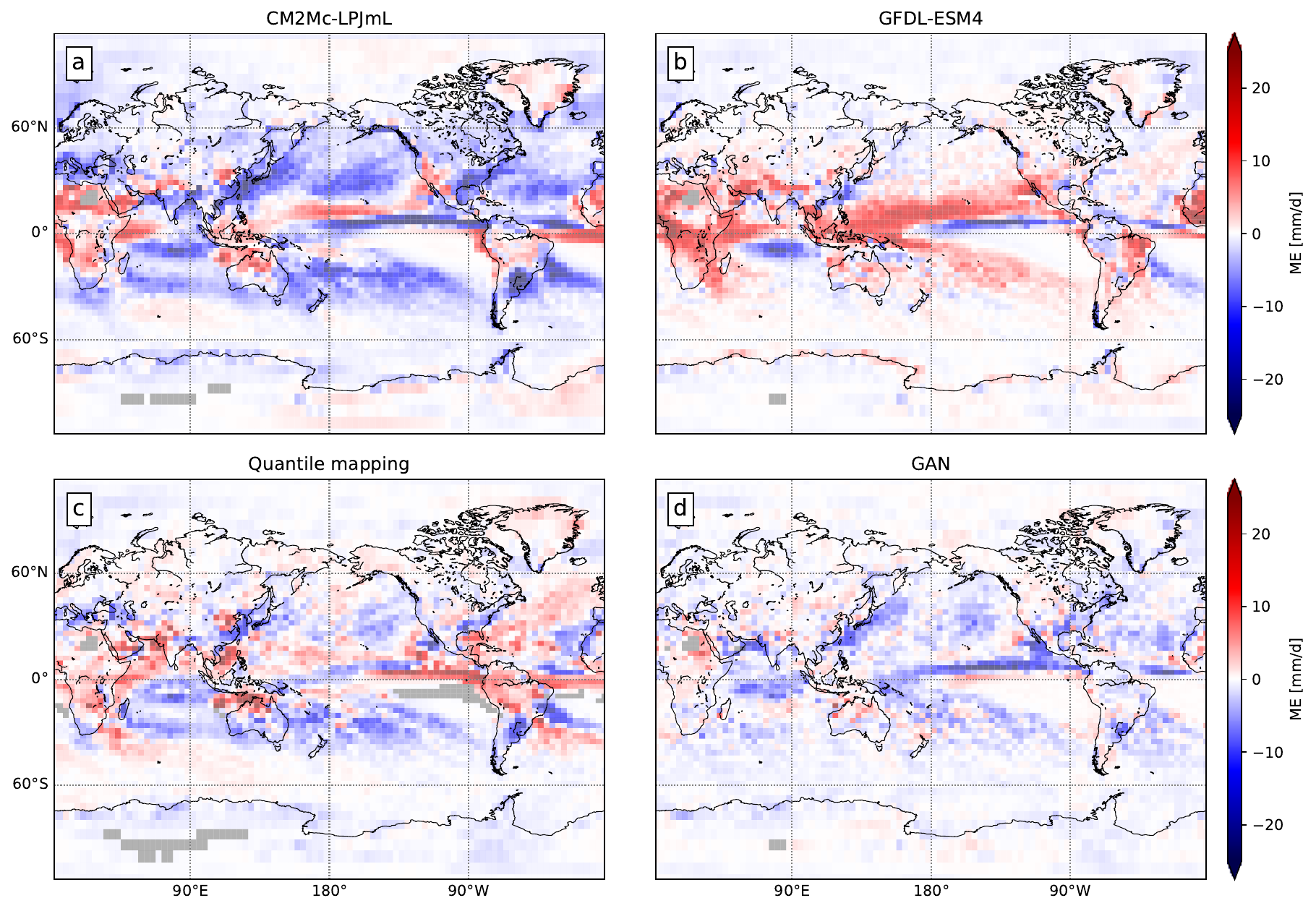}
    \caption{Global maps showing the difference in the 95th precipitation percentile for the SON season of the test set. For (a) CM2Mc-LPJmL, (b) GFDL-ESM4, (c) QM-based and (d) GAN-based post-processing methods applied to the CM2Mc-LPJmL output. Grid cells where the percentiles could not be determined due to insufficient statistics are shown in grey.}
\label{fig:son_extremes_bias}
\end{figure} 


\begin{figure}[ht]
    \centering
    \includegraphics[width=\textwidth]{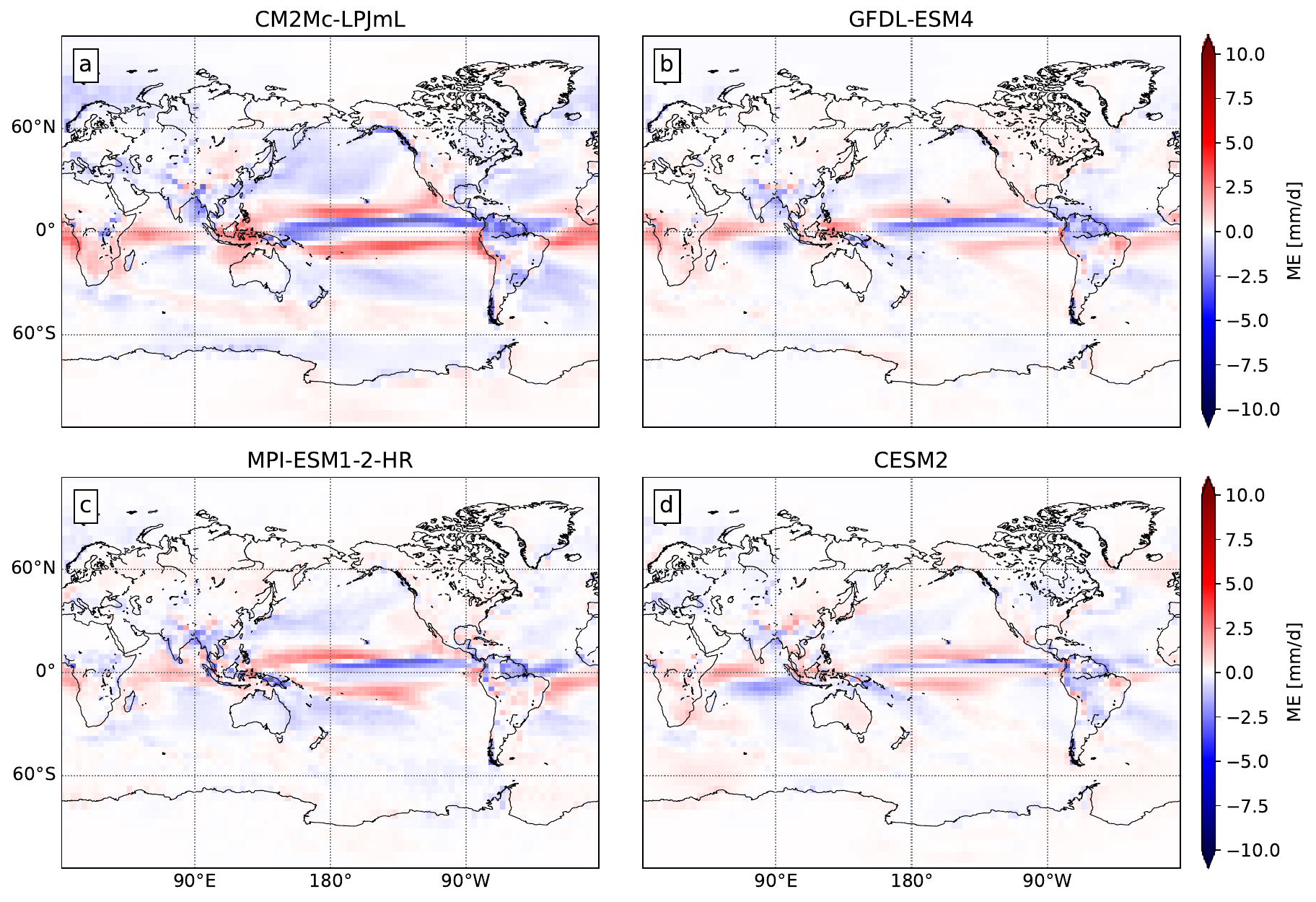}
    \caption{Global maps showing the mean error for the annual time series of the test set. For (a) CM2Mc-LPJmL, (b) GFDL-ESM4, (c) MPI-ESM1-2-HR and (d) CESM2.}
\label{fig:cmip_bias}
\end{figure}


\begin{figure}[ht]
    \centering
    \includegraphics[width=\textwidth]{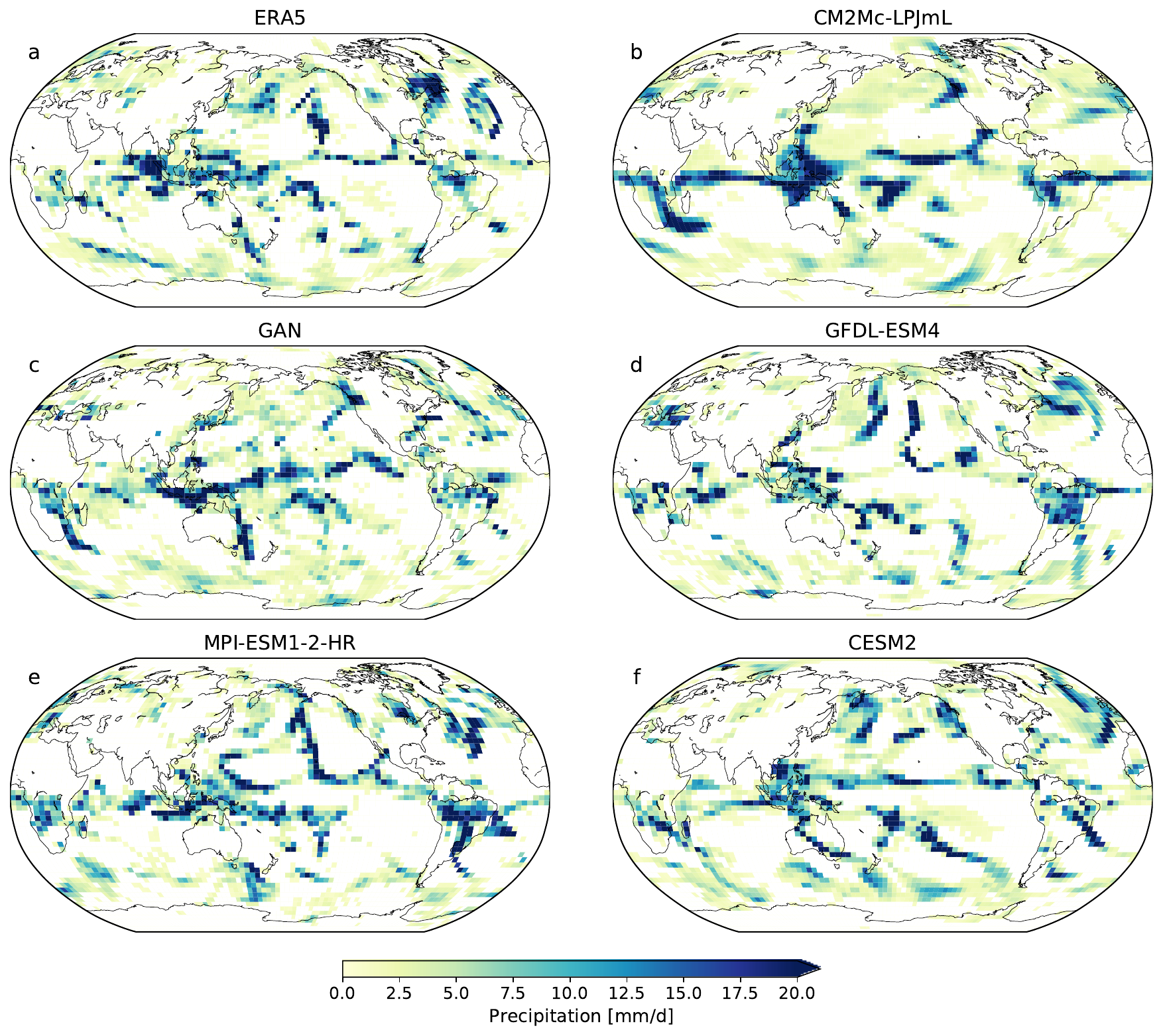}
    \caption{Qualitative and quantitaive comparison of the intermittency in daily precipitation above 1~mm/day, on the same date (25th December 2014), for the (a) ERA5 reanalysis, (b) CM2Mc-LPJmL model, (c) GAN-based post-processing, (d) GFDL-ESM4, (e) MPI-ESM1-1-HR and (f) CESM2.}
\label{fig:cmip_single_frame}
\end{figure} 

\begin{figure}[ht]
    \centering
    \includegraphics[width=0.6\textwidth]{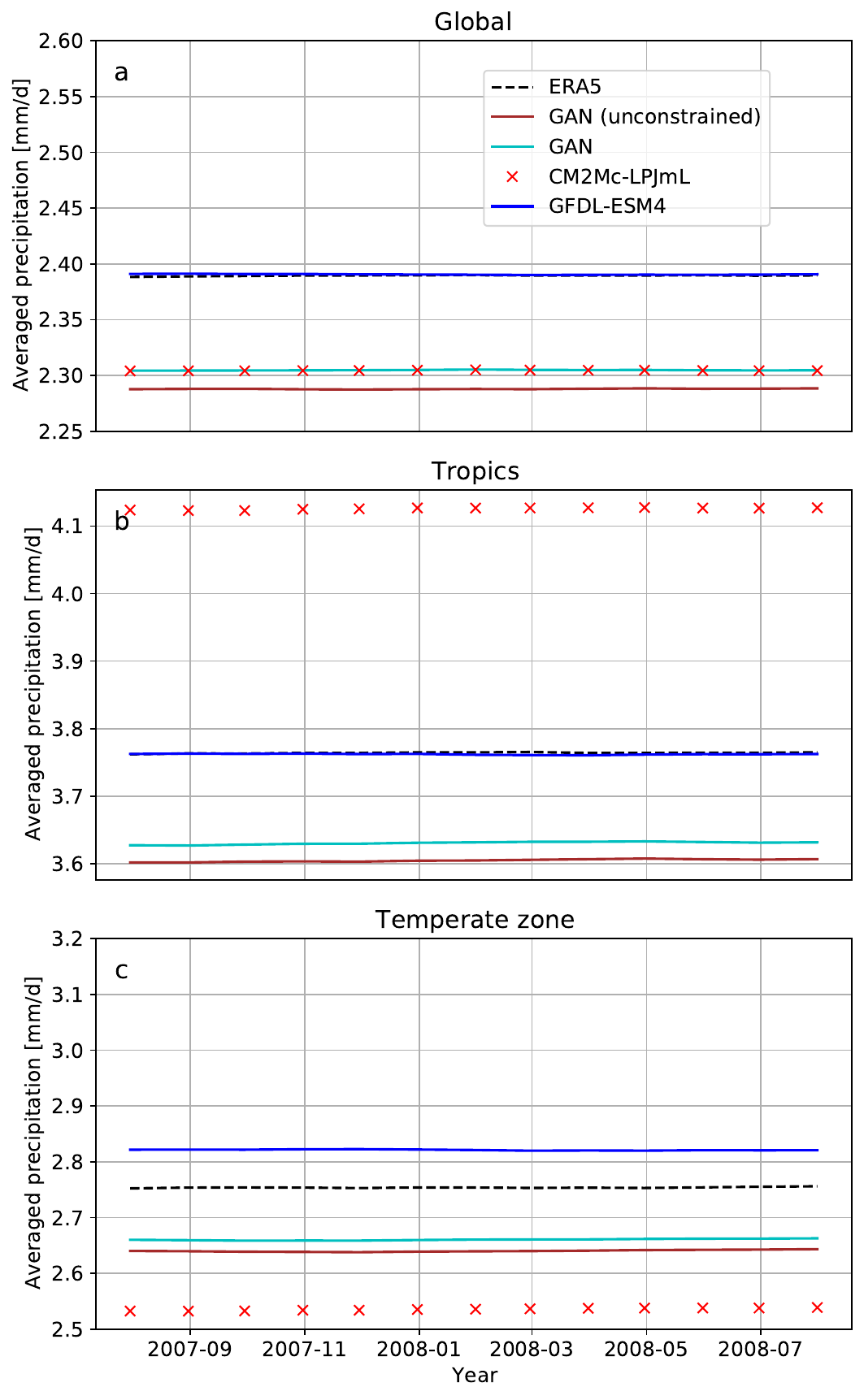}
    \caption{ Large-scale trends as a three year rolling-mean of monthly and spatially average precipitation for the test set period.
    For (a) global data, (b) the tropics and (c) temperate zone, of the ERA5 reanalysis (black dotted line), CM2Mc-LPJmL (red crosses) and GFLD-ESM4 (blue) models, as well as the constrained (cyan) and unconstrained (brown) GANs.}
    
\label{fig:historical_time_series}

\end{figure} 
\begin{figure}[ht]
    \centering
    \includegraphics[width=0.6\textwidth]{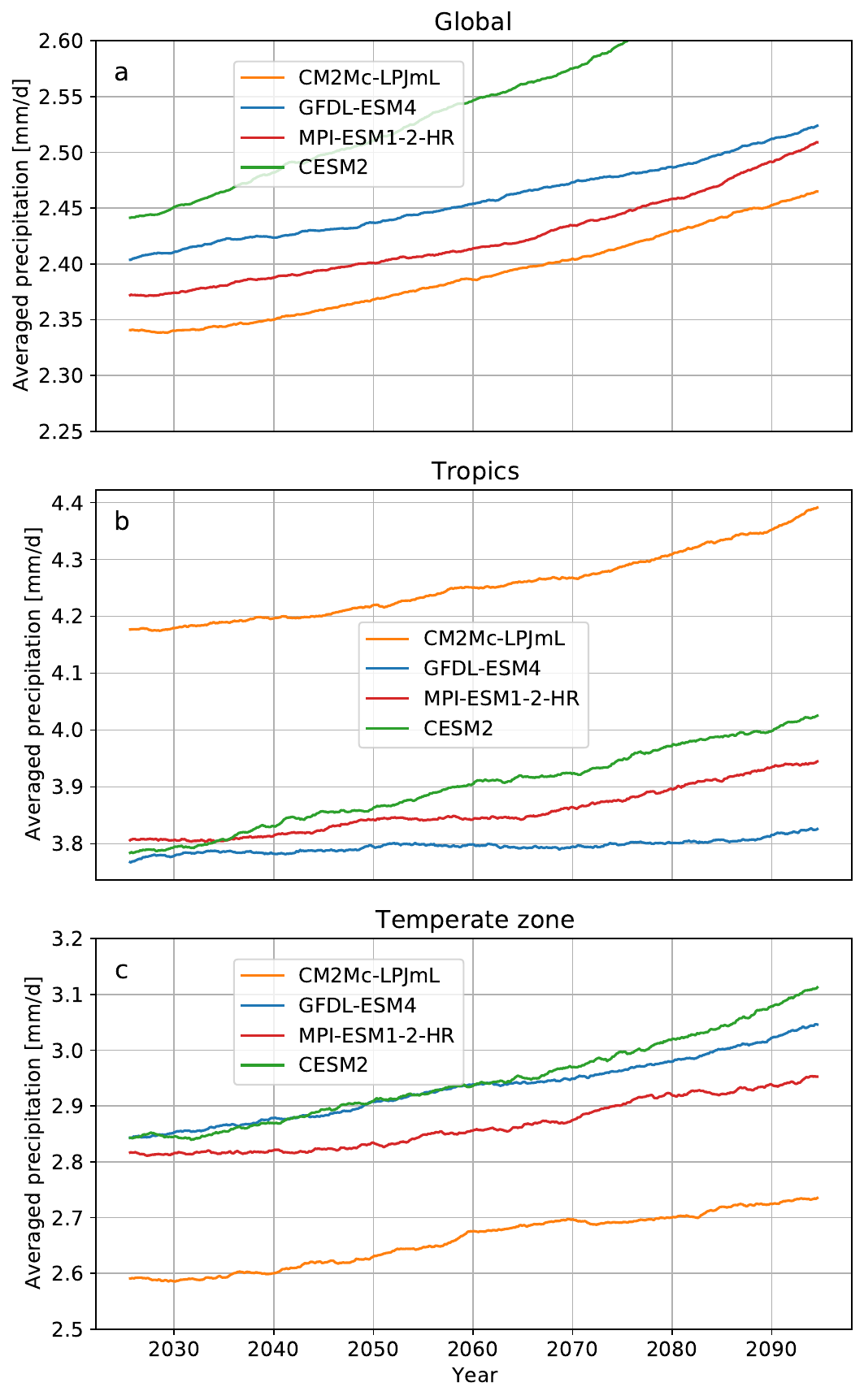}
    \caption{Large-scale precipitation trends are shown for the CMIP6 SSP5-8.5 scenario for the global time series (a), the tropics and temperate zone (c), of the CM2Mc-LPJmL (orange), GFLD-ESM4 (blue), MPI-ESM1-1-HR (red) and CESM2 (green) model.}
\label{fig:cmip_non_stationarity}
\end{figure} 

\begin{figure}[ht]
    \centering
    \includegraphics[width=1.0\textwidth]{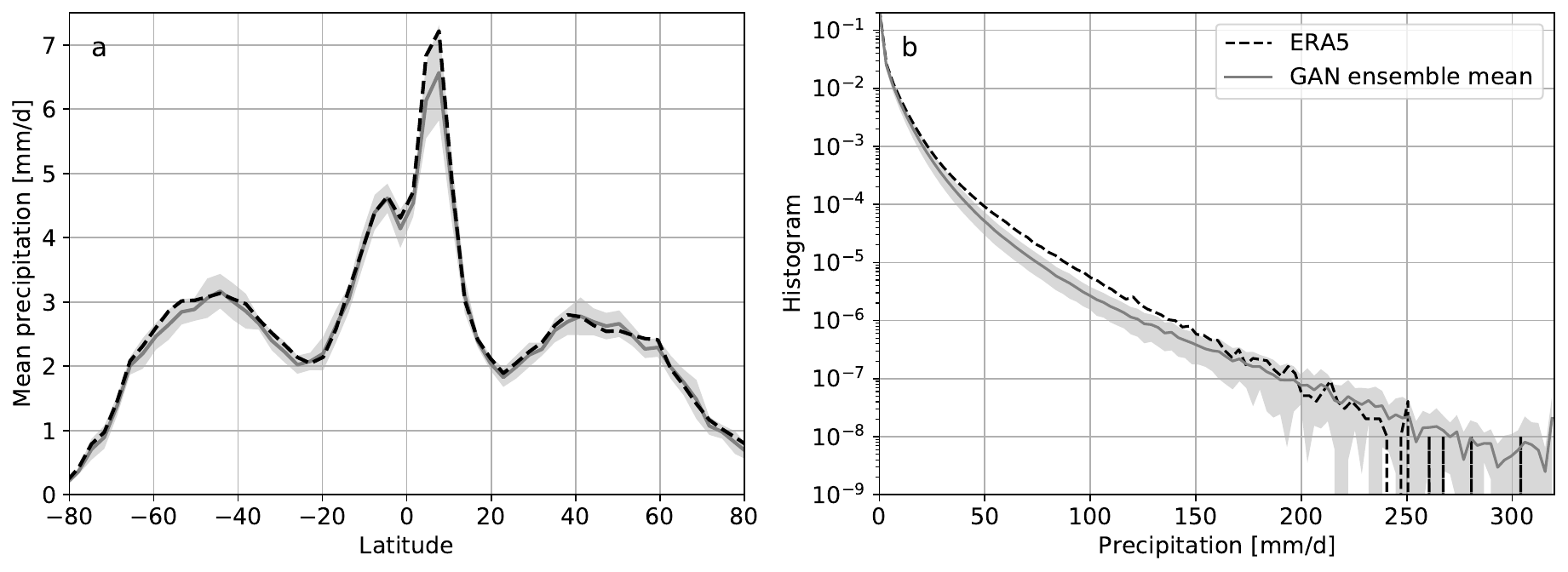}
    \caption{Long-term precipitation statistics based on latitude-profiles and relative frequency histograms for the ERA5 reanalysis (black dotted line) and the ensemble mean of ten GANs (grey, standard deviation as shades) with the same hyperparameters but different checkpoints during the training. }
\label{fig:gan_ensemble}
\end{figure} 

\clearpage
\begin{table}[ht]
            \centering
            \caption{
                The averaged absolute value of the grid-cell wise bias is shown for the raw model output of CM2Mc-LPJmL, GFDL-ESM4, MPI-ESM1-1-HR and CESM2.}
            \begin{tabular}{lcccccccc}
               \hline
               Season   & CM2Mc-LPJmL  & GFDL-ESM4 &   MPI-ESM1-2-HR   & CESM2  \\ 
               \hline
               Annual   & 0.769        & 0.448     &   0.516           & 0.404  \\
               DJF      & 0.915        & 0.544     &   0.677           & 0.530  \\
               MAM      & 0.886        & 0.603     &   0.702           & 0.549  \\
               JJA      & 0.963        & 0.589     &   0.649           & 0.584  \\
               SON      & 0.823        & 0.508     &   0.595           & 0.513  \\
               \hline
            \label{tab:bias}
            \end{tabular}       
\end{table} 

\begin{table}[h!]
            \centering
            \caption{
                The averaged absolute error of the grid-cell-wise 95th precipitation percentiles for the raw CM2Mc-LPJmL and GFDL-ESM4 models, as well as for the QM- and GAN-based post-processing, using the CM2Mc-LPJmL output as input.}
            \begin{tabular}{lcccccccc}
               \hline
               Season   & CM2Mc-LPJmL  & GFDL-ESM4 & \%   & QM    & \%   & GAN            & \% \\ 
               \hline
               Annual   & 3.715        & 2.774     & 25.33  & 1.868 &  49.72 & \textbf{1.495}  & \textbf{59.76} \\
               DJF      & 4.198        & 3.071     & 26.85  & 3.480 &  17.10 & \textbf{1.889}  & \textbf{55.63}\\
               MAM      & 4.200        & 3.114     & 25.86  & 2.954 &  29.67 & \textbf{1.876}  & \textbf{55.34}\\
               JJA      & 4.324        & 2.995     & 30.73  & 3.077 &  28.84 & \textbf{1.889}  & \textbf{56.31} \\
               SON      & 3.875        & 2.826     & 27.07  & 2.818 &  27.28 & \textbf{1.972}  & \textbf{49.11} \\
               \hline
            \label{tab:bias95}
            \end{tabular}       
\end{table}